%% file: sam.tex
\documentclass{article}

\input preamble.tex

\def\sscoin{%
  \leavevmode
  \vtop{\offinterlineskip 
    \setbox0=\hbox{\scriptsize S}%
    \setbox2=\hbox to\wd0{\hfil\hskip-.03em
    \vrule height .3ex width .08ex\hskip .08em
    \vrule height .3ex width .08ex\hfil}
    \vbox{\copy2\box0}\box2}}
\newcommand\affil[2]{%
  \begingroup
  \renewcommand\thefootnote{}\footnote{\llap{$\hbox{}^{#1}\hbox{}$}#2}%
  \addtocounter{footnote}{-1}%
  \endgroup
}
\newcommand\markonly[1]{%
$\hbox{}^{\mbox{\kern4.5pt,\kern0.75pt #1}}$
}

\advance\oddsidemargin by -0.525in
\advance\textwidth by 1.05in
\advance\topmargin by -0.625in
\advance\textheight by 1.25in

\title{\vspace{-0.5in}Cluster Analysis of Malware Family Relationships}

\author{
Samanvitha Basole\thanks{s97basole@gmail.com}\ \ \ \  
Mark Stamp\thanks{mark.stamp@sjsu.edu}\markonly{\sscoin}
}

\date{}

\begin{document}

\maketitle

\vglue-0.35in

\affil{\sscoin}{Department 
of Computer Science,
San Jose State University,
San Jose, California}

\abstract
In this paper, we use $K$-means clustering to analyze various relationships 
between malware samples. We consider a dataset comprising~20 malware families
with~1000 samples per family. These families can be categorized into seven different
types of malware. We perform clustering based on pairs of families and use the
results to determine relationships between families. We perform a similar cluster
analysis based on malware type. Our results indicate that $K$-means clustering
can be a powerful tool for data exploration of malware family relationships.

\section{Introduction}

Previous research has demonstrated that it is possible in some cases
to train a machine learning model to detect multiple malware families~\cite{basole2020multifamily}. 
Specifically, neighborhood-based techniques are relatively effective in such a situation. 
Although support vector machines (SVM) did not perform well in this previous research,
both $k$-nearest neighbors ($k$-NN) and random forest (RF) were able to distinguish malware 
from benign with good accuracy, even when several malware families were combined 
to form the malware class.

In this research, we consider the same dataset used in~\cite{basole2020multifamily},
which includes~20 malware families. Here, we apply cluster analysis to these families. 
Our goal is to determine whether we can discover interesting connections, relationships, 
and differences between these various families, based on elementary clustering techniques. 
While~\cite{basole2020multifamily} considers binary classification experiments to 
distinguish malware from benign, the research in this paper is focused on a
data exploration problem. The features we use are byte $n$-gram frequencies, 
while the clustering method we consider is the well-known $K$-means algorithm.

The remainder of this paper is organized as follows. In Section~\ref{sec:bac}, we provide
relevant background information, including a brief discussion of related work.
Then in Section~\ref{sec:exp}, we give our experimental results and analysis. 
Finally, Section~\ref{sec:con} concludes this paper, where we have included
suggestions for future work.

\section{Background}\label{sec:bac}

In this section, we first consider relevant related work. Then we discuss our malware dataset,
and we provide information on each family in the dataset. 
We also present the various metrics that we use in our
clustering experiments. Finally, we provide an introduction to
the $K$-means clustering algorithm.

\subsection{Related Work}

In~\cite{pirscoveanu2015analysis}, the authors use API calls to classify malware based on their 
types. They consider a random forest (RF) classifier and achieve an average 
area under the ROC curve (AUC) of~0.98. In contrast, we use $n$-grams and 
a clustering approach, and we are considering a data exploration problem,
rather than a straightforward classification problem.
 
The authors in~\cite{bayer2009scalable} propose a clustering approach for malware, 
wherein the goal is to cluster samples based on similar behavior. These authors use 
features such as system calls and network activity, to cluster malicious 
code based on its behavior. In contrast to this previous work, our cluster analysis
is based on simpler and easier to collect features and, again, we are in a
data exploration mode of operation.

The research in~\cite{pitolli2017malware} uses BIRCH clustering based on 
static and dynamic features. These authors consider~18 families, but~12 of those families 
contain less than~100 samples each. Our approach uses a larger and balanced 
dataset for clustering. 

\subsection{Dataset}

The same dataset as used in~\cite{basole2020multifamily} is considered in
this research. This dataset includes~20 
families, which we categorize into malware types, 
as listed in Table~\ref{tab:families}. 


\begin{table}[!htb]
\begin{center}
\caption{Type of each malware family}\label{tab:families}
{\footnotesize
\begin{tabular}{ccc|ccc}\midrule\midrule
\textbf{Index} & \textbf{Family} & \textbf{Type} & \textbf{Index} & \textbf{Family}  & \textbf{Type} \\
\midrule
0 & Adload~\cite{adload}          & Trojan Downloader & 10 & Obfuscator~\cite{obfuscator}          & VirTool \\
1 & Agent~\cite{agent}              & Trojan                     & 11 & OnLineGames~\cite{onlinegames} & Password Stealer \\
2 & Alureon~\cite{alureon}        & Trojan                     & 12 & Rbot~\cite{rbot}                               & Backdoor \\
3 & BHO~\cite{bho}                  & Trojan                      & 13 & Renos~\cite{renos}                      & Trojan Downloader \\
4 & CeeInject~\cite{ceeinject}  & VirTool                     & 14 & Startpage~\cite{startpage}              & Trojan \\
5 & Cycbot.G~\cite{cycbot}      & Backdoor                & 15 & Vobfus~\cite{vobfus}                        & Worm \\
6 & DelfInject~\cite{delfinject}  & VirTool                     & 16 & Vundo~\cite{vundo}                      & Trojan Downloader \\
7 & FakeRean~\cite{fakerean} & Rogue                     & 17 & Winwebsec~\cite{winwebsec}        & Rogue \\
8 & Hotbar~\cite{hotbar}          & Adware                    & 18 & Zbot~\cite{zbot}                              & Password Stealer \\
9 & Lolyda.BF~\cite{lolyda}     & Password Stealer    & 19 & Zeroaccess~\cite{zeroaccess}       & Trojan \\
\midrule\midrule 
\end{tabular}
}
\end{center}
\end{table}

Each of the malware families in Table~\ref{tab:families} is summarized below.
\begin{description}
\item[\bf Adload] downloads an executable file, stores it remotely, executes the file, and disables 
proxy settings~\cite{adload}. 
\item[\bf Agent] downloads Trojans or other software from a remote server~\cite{agent}. 
\item[\bf Alureon] exfiltrates usernames, passwords, credit card data, and 
other confidential data from an infected system~\cite{alureon}. 
\item[\bf BHO] can perform a variety of actions, guided by an attacker~\cite{bho}. 
\item[\bf CeeInject] uses advanced obfuscation to avoid being detected by antivirus software~\cite{ceeinject}. 
\item[\bf Cycbot.G] connects to a remote server, exploits vulnerabilities, and spreads through backdoor 
ports~\cite{cycbot}. 
\item[\bf DelfInject] sends usernames, passwords, and other personal 
and private information to an attacker~\cite{delfinject}. 
\item[\bf FakeRean] pretends to scan the system, notifies the user of supposed issues, 
and asks the user to pay to clean the system~\cite{fakerean}. 
\item[\bf Hotbar] is adware that shows ads on webpages and installs additional adware~\cite{hotbar}. 
\item[\bf Lolyda.BF] sends information from an infected computer and monitors the system. It can share user 
credentials and network activity with an attacker~\cite{lolyda}. 
\item[\bf Obfuscator] tries to obfuscate or hide itself to defeat malware detectors~\cite{obfuscator}.
\item[\bf OnLineGames] steals login information of online games and tracks user 
keystroke activity~\cite{onlinegames}. 
\item[\bf Rbot] gives control to attackers via a backdoor that can be used to access information or
launch attacks, and serves as a gateway to infect additional sites~\cite{rbot}.
\item[\bf Renos] downloads software that claims the system has spyware and asks for a payment to 
remove the nonexistent spyware~\cite{renos}. 
\item[\bf Startpage] changes the default browser homepage and may perform other
malicious activities~\cite{startpage}. 
\item[\bf Vobfus] is a worm that downloads malware and spreads through USB drives or other 
removable devices~\cite{vobfus}. 
\item[\bf Vundo] displays pop-up ads and may download files. It uses advanced techniques to 
defeat detection~\cite{vundo}.
\item[\bf Winwebsec] displays alerts that ask the user for money to 
fix supposed issues~\cite{winwebsec}.
\item[\bf Zbot] is installed through email and shares a user's personal information with attackers.
In addition, Zbot can disable a firewall~\cite{zbot}.
\item[\bf Zeroaccess] is a Trojan horse that downloads applications that click on ads,
thereby making money for the attacker~\cite{zeroaccess}.
\end{description}

The features we use for clustering are based on byte $n$-gram frequencies. 
Specifically, we choose the 
top~20 byte $n$-grams, with~$n=2$, from our malware class. 
These frequency vectors are then normalized, 
so that each vector can be viewed as a discrete probability distribution.
The resulting normalized bigram frequency
vectors (of length~20) form our feature set. 

Our experiments include clustering based on pairs of families, clustering selected families belonging
to different malware types, and clustering families belonging to the same malware type.  
The number of samples of each of the seven malware types found in our dataset is given in
Table~\ref{tab:malwareTypes}.
Note that we categorize ``Trojan Downloader'' as a type of Trojan, giving 
us the seven distinct types listed in Table~\ref{tab:malwareTypes}.

\begin{table}[!htb]
\begin{center}
\caption{Number of samples of each type}\label{tab:malwareTypes}
\begin{tabular}{cc}\midrule\midrule
\textbf{Malware Type} & \textbf{Samples} \\
\midrule
VirTool        &  3000    \\
Password Stealer       &     3000\\
Backdoor       &    2000 \\
Trojan          &   8000 \\
Worm      & 1000      \\
Rogue       &    2000           \\
Adware          &       1000   \\
\midrule\midrule 
\end{tabular}
\end{center}
\end{table}

\subsection{Metrics}

In this section, we discuss the metrics used to numerically evaluate our clustering results. 
Note that we do not use accuracy, due to the label-switching problem that occurs
when we attempt to apply this metric to clustering results~\cite{stephens2000dealing}. 

One popular choice for clustering is the so-called $v$-measure, which is a robust metric for 
cluster evaluation---robust in the sense that a permutation of the cluster labels does not affect 
the score. The $v$-measure is defined as the harmonic mean between homogeneity 
(i.e., the case where each cluster contains all points from a single class) and completeness 
(i.e., the case where all points from the same class are in one cluster)~\cite{scikit-learn}. 
Another nice feature of this metric is that it is symmetric, in 
that it yields the same score if the predicted classes 
and the true classes are switched. The $v$-measure ranges from~0 to~1. 

Although $v$-measure is a robust evaluation metric with many advantages, it is not the best choice 
for the research considered in this paper. The $v$-measure is not normalized for random 
cluster results, hence it would tend to produce a higher score for random cluster assignments when 
a ``large'' number of clusters~$K$ is chosen, say, $K > 10$.
In contrast, adjusted mutual information (AMI) 
results in random label assignments having a score close to~0, 
regardless of the size of the dataset or the number of clusters.

Another useful metric for clustering is the adjusted Rand index (ARI). 
The ARI is similar to AMI, in the sense that it is adjusted to account for chance. 
The Rand index is a similarity measure that considers all pairs of 
samples and uses the number of pairwise agreements in the true and predicted clusters. 
Specifically, the Rand index is calculated as~\cite{santos2009use}
$$ 
  \text{RI} = \dfrac{a + d}{a + b + c + d} = \frac{a + d}{\displaystyle{n \choose 2}}
$$
where~$a$, $b$, $c$, and~$d$ are defined as follows:
If~$U$ and~$V$ are two different partitions or clusterings of the same data, 
then let~$a$, $b$, $c$, and~$d$ be the number of objects determined by
\begin{align*}
a &= \mbox{in the same cluster in $U$ and in the same cluster in $V$}\\
b &= \mbox{in the same cluster in $U$ but in different clusters in $V$}\\
c &= \mbox{in the same cluster in $V$ but in different clusters in $U$}\\
d &= \mbox{in different clusters in $U$ and in different clusters in $V$}
\end{align*}
The formula for ARI is calculated using the raw Rand index~RI as~\cite{scikit-learn}
$$ 
  \mbox{ARI} 
      = \frac{\RI - \mbox{E}(\RI)}{
          \max(\RI) - \mbox{E}(\RI)}
$$
where~$E$ is the expected value operator.

The authors in~\cite{romano2016adjusting} state that AMI should be used when 
the true clusters are unbalanced in size, while ARI should be used when the 
true clusters are large and roughly equal-sized. In our research, the size of the 
ground truth for family labels is precisely balanced with~1000 samples in each family. 
Thus, we use ARI to evaluate our clustering predictions.

\subsection{$K$-Means}

In this section,
we first discuss a generic approach to clustering.
We then consider how to implement such an approach, which
leads directly to the $K$-means algorithm.

Suppose that we are given the~$n$ data points
$
  X_1, X_2, \ldots, X_n
$,
where each of the~$X_i$ is a vector of~$m$ real numbers. 
For example, we could analyze a set of malware
samples based on, say, 
five distinct scores, denoted~$s_1,s_2,\ldots,s_5$. 
Then each data point would be of the form
$$
  X_i = (s_1, s_2, s_3, \ldots, s_5) .
$$

We assume that the desired number of clusters~$K$ is specified in advance
and that we want to partition our~$n$ data points~$X_1,X_2,\ldots,X_n$ 
into~$K$ clusters. 
We also assume that we have a distance function~$d(X_i,X_j)$
that is defined for all pairs of data points.

We associate a centroid
with each cluster, where the centroid can be viewed as representative 
of its cluster. Intuitively, a centroid is the center of mass of a cluster.
We denote cluster~$j$ as~$C_j$ with the 
corresponding centroid denoted as~$c_j$.
Note that in $K$-means, centroids need not be actual data points.

Now, suppose that we have clustered our~$n$ data points. 
Then we have a set of~$K$ centroids,
$$
  c_1, c_2, c_3, \ldots, c_K 
$$
and each data point is associated with exactly one centroid.
Let~$\centroid{X_i}$ denote the (unique) centroid of the cluster
to which~$X_i$ belongs. 
The centroids determine the clusters,
in the sense that whenever we have
$$
  c_j = \centroid{X_i} ,
$$
then~$X_i$ belongs to cluster~$C_j$. 

Before we can cluster data based on the process outlined above, 
we need to address the following two questions.
\begin{enumerate}
\item How do we determine the centroids~$c_j$?
\item How do we determine the clusters? That is,
we need to specify the function~$\centroid{X_i}$,
which assigns data points to centroids. This has the effect
of determining the clusters.
\end{enumerate}
There are many reasonable ways to answer these questions. 
Next, we consider an intuitively appealing approach that
leads directly to the $K$-means algorithm.

Intuitively, it seems clear that the more ``compact'' a cluster is, 
the better. Of course, this will depend on the data points~$X_i$ and 
the number of clusters~$K$. Since the data is given,
and we are assuming that~$K$ has been specified,
we have no control over the~$X_i$ or~$K$.  But, we do have
control over the selection of the centroids~$c_j$ and the assignment 
of points to centroids via the function~$\centroid{X_i}$. 
The choice of centroids and the assignment of
points to centroids will clearly influence the compactness
(i.e., ``shape'') of the resulting clusters. 

Let
\begin{equation}\label{eq:cluster_distortion}
  \distortion{} = \sum_{i=1}^n d\bigl(X_i,\centroid{X_i}\bigr) .
\end{equation}
Intuitively, the smaller the~$\distortion{}$, the better, since a smaller~$\distortion{}$
implies that individual clusters are more compact.\footnote{In addition to having 
compact clusters, we might also want a large separation between clusters. 
However, such separation is not (directly) accounted for in $K$-means.}

\begin{figure}[!htb]
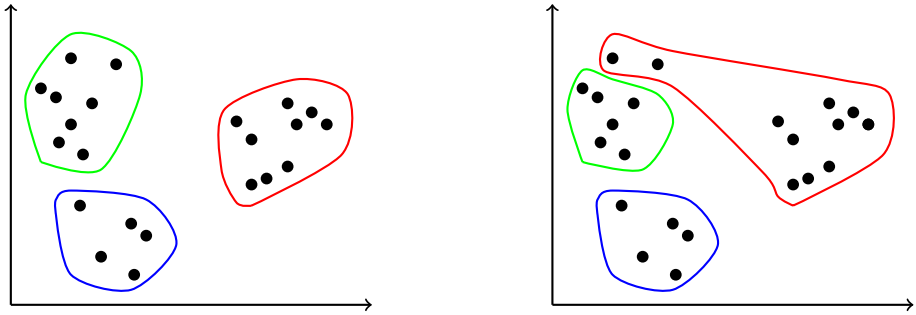

  \centering
    \input figures/figure_3.tex
  \caption{Smaller and larger $\distortion{}$ for the same dataset}\label{fig:clustDistortion}
\end{figure}

For example, consider the data in Figure~\ref{fig:clustDistortion}, where the same
data points are clustered in two different ways.  
It is clear that the clustering on 
the left-hand side in Figure~\ref{fig:clustDistortion} has a smaller $\distortion{}$ than 
that on the right-hand side. Therefore, we would say that the left-hand clustering is superior,
at least with respect to the measure of~$\distortion{}$.

Suppose that we try to minimize the $\distortion{}$. 
First, we observe that~$\distortion{}$ depends on~$K$, 
since more clusters implies more centroids---in the limit, we
could let~$K=n$, and make each data point a centroid,
in which case the~$\distortion{}$ is~0. To emphasize this
dependence on~$K$, 
we write $\distortion{K}$. As mentioned above, 
we assume that~$K$ is specified in advance.

The problem we want to solve can be stated precisely as
\begin{equation}\label{eq:clustProb} 
\begin{split}
  & \mbox{Given: } K \mbox{ and data points } X_1,X_2,\ldots, X_n\\
  & \mbox{Minimize: } \distortion{K} = \sum_{i=1}^n d\bigl(X_i,\centroid{X_i}\bigr) .
\end{split}
\end{equation}
Finding an exact solution to this problem is computationally infeasible.
But, there is a simple iterative approximation that works well in practice.

We claim that a solution to~\eref{eq:clustProb} must satisfy the following two conditions.
\begin{description}
\item[\bf Condition~1] Each~$X_i$ is clustered according to its nearest centroid. That is,
if the data point~$X_i$ belongs to cluster~$C_j$, 
then~$d(X_i,c_j) \leq d(X_i,c_{\ell})$ for all~$\ell\in\{1,2,\ldots,K\}$,
where the~$c_{\ell}$ are the centroids.
\item[\bf Condition~2] Each centroid is located at the center of its cluster.
\end{description}

To verify the necessity of Condition~1,
suppose that~$X_i$ is in cluster~$C_j$ and that we have~$d(X_i,c_{\ell}) < d(X_i,c_{j})$ 
for some~$\ell$. Then by simply reassigning~$X_i$ to cluster~$\ell$, we 
will reduce~$\distortion{K}$.
Condition~2 also seems intuitively clear, and 
it is a straightforward calculus exercise to prove the necessity 
of this condition as well.

Condition~1 tells us that given any clustering for which there are 
points that are not assigned to
their nearest centroid, we can improve the clustering by simply reassigning
such data points to their nearest centroid. Condition~2 implies that
we always want a centroid to be at the center of its cluster. Therefore, given any
clustering, we may improve it---and we cannot make it worse---with 
respect to~$\distortion{K}$ by performing 
either of the following two steps.

\begin{description}
\item[\bf Step~1] Assign each data point to its nearest centroid.
\item[\bf Step~2] Recompute the centroids so that each lies at the center of
its respective cluster.
\end{description}

It is clear that nothing can be gained by applying 
Step~1 more than once in succession, and the same holds true for Step~2.
However, by alternating between these two steps, we obtain an iterative process that
yields a series of solutions that will generally tend to improve, and can
never get worse with respect to~$\distortion{K}$. 
This is precisely the $K$-means algorithm~\cite{clustMoore},
which we state somewhat more precisely in Table~\ref{alg:clustKmeans}.

\begin{table}
\caption{$K$-means clustering}\label{alg:clustKmeans}
\centering
\begin{minipage}{0.9\linewidth}
\begin{algorithmic}[1]
\Given{Data points~$X_1,X_2,\ldots,X_n$ to cluster\\
Number of clusters~$K_{\vphantom{M_{\sum}}}$}
\Initialize{Partition~$X_1,X_2,\ldots,X_n$ into clusters~$C_1,C_2,\ldots,C_K$}
\While{stopping criteria is not met}
\For{$j=1$ to $K$}
\State Let centroid~$c_j$ be the center of cluster~$C_j$
\EndFor
\For{$i = 1$ to $n$}
\State Assign~$X_i$ to cluster~$C_j$ so that~$d(X_i,c_j)\leq d(X_i,c_{\ell})$
\Statex \hspace*{0.7in} for all $\ell\in\{1,2,\ldots,K\}$
\EndFor
\EndWhile
\end{algorithmic}
\end{minipage}
\end{table}

The stopping criteria in Table~\ref{alg:clustKmeans} 
could be that~$\distortion{K}$ improves (i.e., decreases) by less 
than a set threshold, or that the centroids
do not change by much, or we could simply 
run the algorithm for a fixed number of iterations.

Note that the algorithm in Table~\ref{alg:clustKmeans} is a hill climb,
and hence $K$-means
is only assured of finding a local maximum. And, as with any hill climb,
the maximum we obtain will depend on our choice for the initial conditions. 
For $K$-means, the initial conditions correspond to the initial selection of centroids.
Therefore, it can be beneficial to repeat the algorithm multiple times with
different initializations of the centroids.

In this experiments below, we use the $K$-means clustering algorithm to explore the 
natural formation of malware clusters. We employ elbow plots as a tool to 
discern structure from the~20 malware families based on pairwise clusters.
Next, we discuss elbow plots in this context.

\subsection{Elbow Plots}

Suppose that we graph the clustering error as a function of the number of clusters, $K$.
Then an ``elbow'' in this graph indicates the point where 
adding another cluster does not significantly improve the clustering 
results~\cite{bholowalia2014ebk}. Such an elbow
can be used to determine the (near) optimal number of clusters.

\begin{figure}[!htb]
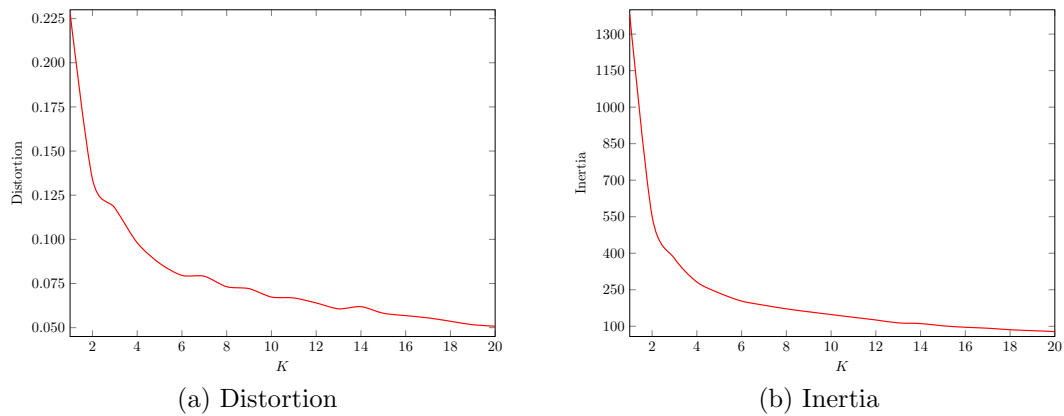

\centering
\begin{tabular}{ccc}
\input figures/elbow_d.tex
& &
\input figures/elbow_i.tex
\\
(a) Distortion
& &
(b) Inertia
\end{tabular}
\caption{Elbow plots}\label{fig:elbowInertDist}
\end{figure}

We choose distortion and inertia for our elbow plots. 
Distortion is calculated as the average of the squared Euclidean distances 
from each point to the nearest centroid, whereas inertia is calculated as 
the sum of these same distances.
For our experiments, elbow plots using distortion and inertia indicate 
that the clusters are not well formed, and thus, the number of clusters is 
somewhat subjective. From Figure~\ref{fig:elbowInertDist}, it appears 
that~$K\in\{4, 5, 6\}$ should be good values for the number of clusters, 
as the inertia and distortion only slightly decrease from that point onward. 
In any case, these elbow plots clearly indicate that the optimal number of
clusters is less than~10, which is somewhat surprising, given that 
we are dealing with~20 families. This is a strong indication that
there is significant similarity between some of the families in our dataset. 

\section{Experiments and Results}\label{sec:exp}

In this section, we present results of three different sets of clustering experiments.
First, we cluster each pair of malware families, and show that we can draw
meaningful conclusions based on these clustering results. Then we consider
clustering experiments where we restrict our attention to one family of each malware type 
under consideration. Finally, we consider clustering multiple families
from the same malware type.

\subsection{Clustering by Family}

In this set of experiments, we perform clustering for each pair of families. 
Since there are~20 families, we have~${20 \choose 2} = 190$ such clustering experiments. 
In each case, the top~20 $n$-grams is extracted to form the features under consideration. 
Every sample in the two families under consideration is then converted to a normalized 
vector of~$n$-gram frequencies. The resulting data is clustered using $K$-means, with~$K=2$. 

\begin{figure}[!htb]
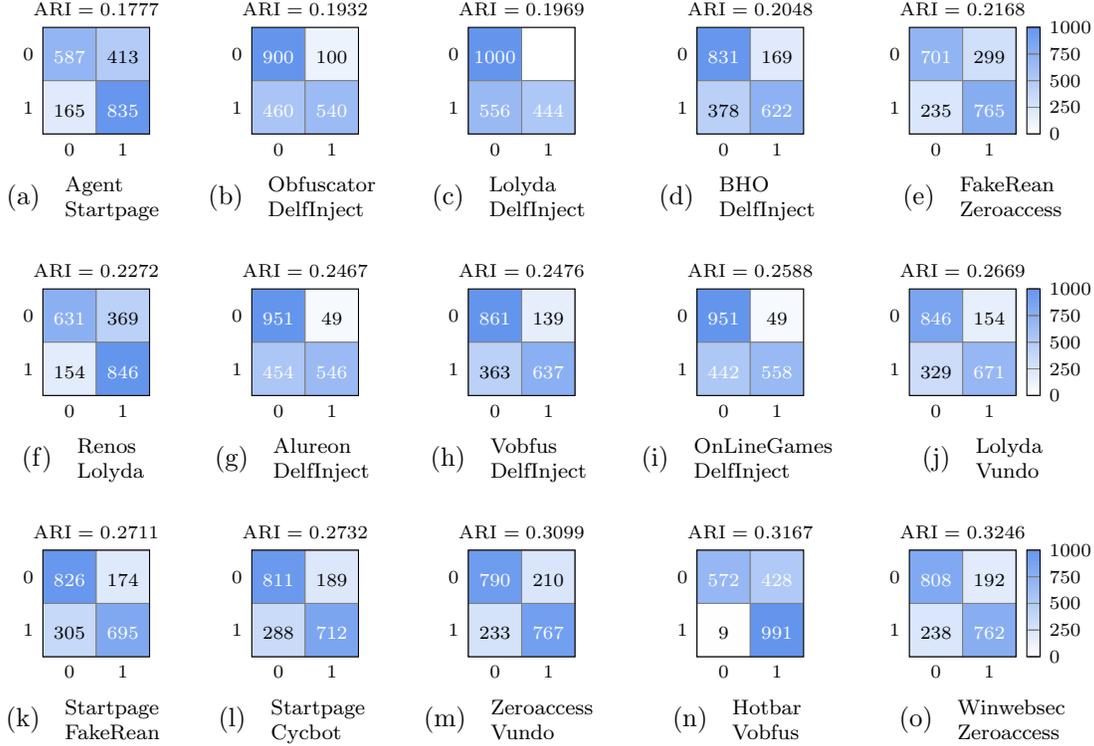

\centering
\begin{tabular}{ccccc}
\input figures/conf2a.tex
& 
\input figures/conf2b.tex
& 
\input figures/conf2c.tex
& 
\input figures/conf2d.tex
& 
\input figures/conf2e_cb.tex
\\
(a) {\footnotesize\begin{tabular}{l} Agent\\ Startpage\end{tabular}}
&  
(b) {\footnotesize\begin{tabular}{l} Obfuscator\\ DelfInject\end{tabular}}
&  
(c) {\footnotesize\begin{tabular}{l} Lolyda\\ DelfInject\end{tabular}}
&  
(d) {\footnotesize\begin{tabular}{l} BHO\\ DelfInject\end{tabular}}
&  
(e) {\footnotesize\begin{tabular}{l} FakeRean\\ Zeroaccess\end{tabular}}
\\
\\
\input figures/conf2f.tex
& 
\input figures/conf2g.tex
& 
\input figures/conf2h.tex
& 
\input figures/conf2i.tex
& 
\input figures/conf2j_cb.tex
\\
(f) {\footnotesize\begin{tabular}{l} Renos\\ Lolyda\end{tabular}}
&  
(g) {\footnotesize\begin{tabular}{l} Alureon\\ DelfInject\end{tabular}}
&  
(h) {\footnotesize\begin{tabular}{l} Vobfus\\ DelfInject\end{tabular}}
&  
(i) {\footnotesize\begin{tabular}{l} OnLineGames\\ DelfInject\end{tabular}}
&  
(j) {\footnotesize\begin{tabular}{l} Lolyda\\ Vundo\end{tabular}}
\\
\\
\input figures/conf2k.tex
& 
\input figures/conf2l.tex
& 
\input figures/conf2m.tex
& 
\input figures/conf2n.tex
& 
\input figures/conf2o_cb.tex
\\
(k) {\footnotesize\begin{tabular}{l} Startpage\\ FakeRean\end{tabular}}
&  
(l) {\footnotesize\begin{tabular}{l} Startpage\\ Cycbot\end{tabular}}
&  
(m) {\footnotesize\begin{tabular}{l} Zeroaccess\\ Vundo\end{tabular}}
&  
(n) {\footnotesize\begin{tabular}{l} Hotbar\\ Vobfus\end{tabular}}
&  
(o) {\footnotesize\begin{tabular}{l} Winwebsec\\ Zeroaccess\end{tabular}}
\end{tabular}
\caption{Selected examples from the~190 pairwise confusion matrices} 
\label{fig:collection}
\end{figure}

The results of these experiments consist of~190 ARI scores 
and~190 confusion matrices. Representative examples of the resulting
confusion matrices are given in Figure~\ref{fig:collection}.  

Each of these~190 clustering experiments provides information on how closely one family is related 
to another. From such results, we can deduce weak and strong links between malware family pairs. 
The~190 ARI similarity scores are given in the form of a heatmap in 
Figure~\ref{fig:hm}. Note that the diagonal elements are~$1$ in every case, since the 
similarity between a family and itself is always~$1$. Also, the heatmap is symmetric,
since the ARI similarity score is itself symmetric. 

\begin{table}[!htb]
  \caption{Heatmap of pairwise clustering ARI scores}\label{fig:hm}
  \centering
    \input figures/heatmap1.tex
\end{table}

\begin{figure}[!htb]
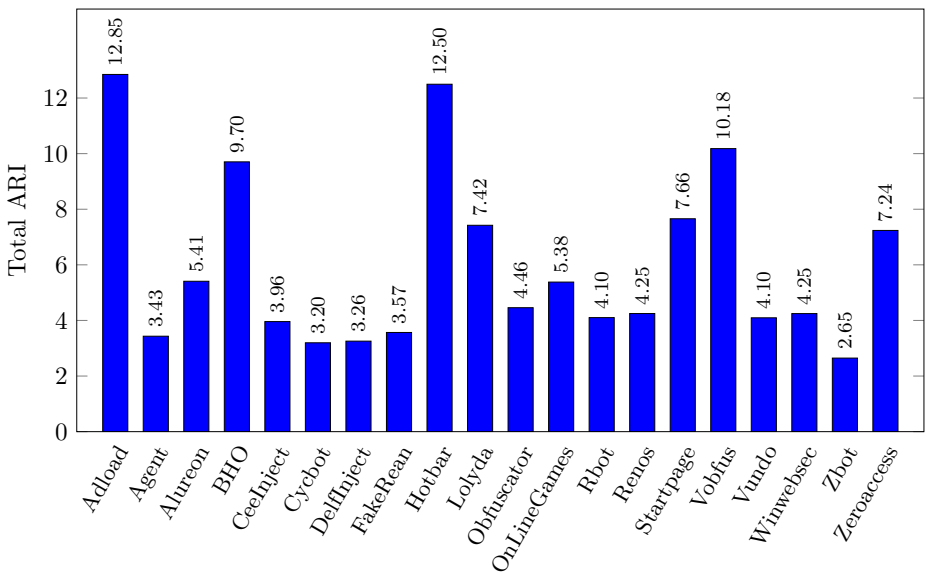

\centering
\input figures/bar_t.tex
\caption{Total ARI score (sum of 19 ARI scores) for each family}\label{fig:totalPairwiseClustering}
\end{figure}

Figure~\ref{fig:totalPairwiseClustering} gives the total pairwise ARI for each family
in the form of a bar graph. That is, each bar represents the sum of the~19 ARI scores
of a given family with all other families in our dataset. We refer to this sum as the total ARI.

\begin{figure}[!htb]
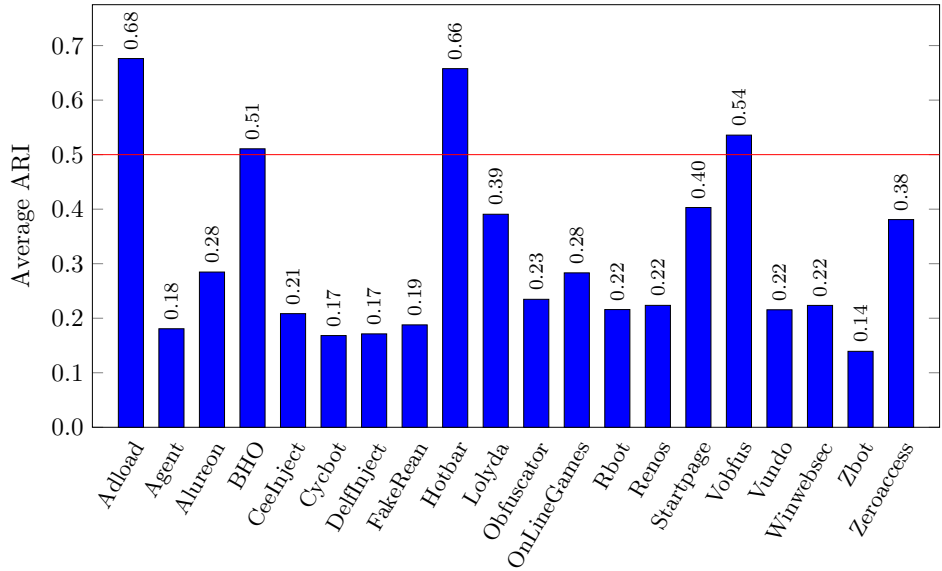

\centering
\input figures/bar_a.tex
\caption{Average ARI score for each family}\label{fig:averagePairwiseClustering}
\end{figure}

In Figure~\ref{fig:averagePairwiseClustering} we give the average ARI for all
pairwise clusters formed with a given family.
Based on the horizontal line at~$y = 0.5$, we see that there are four 
families with a high average ARI, that is, an ARI that exceeds the~$y = 0.5$ line. 
This implies that when each of these four families is clustered against the other families, the ARI is,
on average, particularly high. The four high-ARI families are BHO, Adload, Hotbar, and Vobfus. 
Note that these strong ARI results are also apparent from the total ARI scores
in Figure~\ref{fig:totalPairwiseClustering} and from the heatmap in Figure~\ref{fig:hm}.

From Table~\ref{tab:families},
we see that Hotbar is the only adware in the dataset and Vobfus is the only worm. 
It is intuitive that these malware families would tend to stand out more from the other
families, due to their being of unique types, and would thus be easier to cluster.
This is clearly indicated by the high ARI results for Hotbar and Vobfus. 
On the other hand, BHO and Adload are both Trojans, which is the most common
type in our dataset. This result indicates that in spite of Adload and BHO being Trojans, 
they contain byte bigram features that are significantly different from the other Trojans
in the dataset, namely, Agent, Alureon, Renos, Startpage, Vundo, and Zeroaccess. 
It is also interesting that Adload and BHO are similar to each other, 
in the sense that their pairwise clustering result is poor,
as can be observed from the heatmap in Figure~\ref{fig:hm}.

\begin{figure}[!htb]
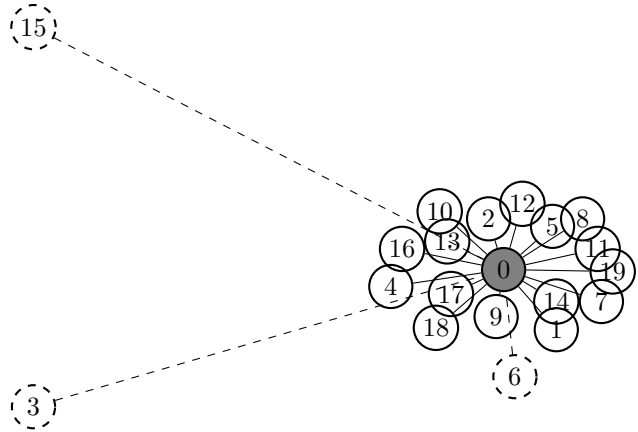

\centering
\input figures/force1c.tex
\caption{Adload relationship with its paired families}\label{fig:adloadPairs}
\end{figure}

\begin{figure}[!htb]
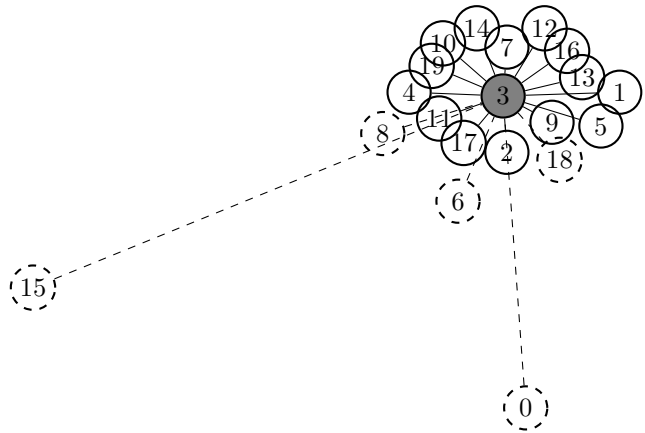

\centering
\input figures/force2c.tex
\caption{BHO relationship with its paired families}\label{fig:BHOPairs}
\end{figure}

\begin{figure}[!htb]
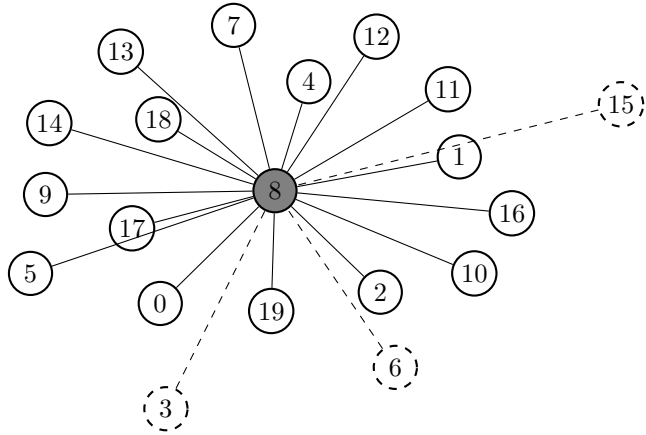

\centering
\input figures/force3c.tex
\caption{Hotbar relationship with its paired families}\label{fig:hotbarPairs}
\end{figure}

\begin{figure}[!htb]
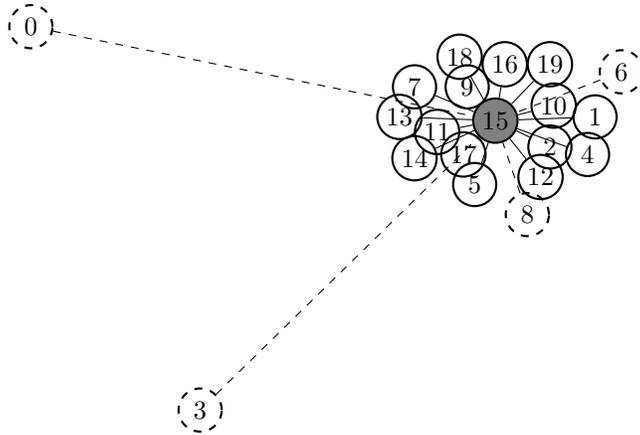

\centering
\input figures/force4c.tex
\caption{Vobfus relationship with its paired families}\label{fig:vobfusPairs}
\end{figure}

To further explore these high ARI families,
we provide graphs showing the relationship strength of each with
respect to all other families. To generate these graphs, we
use the NetworkX library in Python. The resulting graphs are 
given in Figures~\ref{fig:adloadPairs} 
through~\ref{fig:vobfusPairs}, where each node represents a family, with
the node numbers corresponding to
the ``index'' column in Table~\ref{tab:families}. In each of these figures,
the darkened node corresponds to the family mentioned in the caption.
Also, a dotted edge between two nodes indicates an ARI score of~0.5 or less, 
while a solid line represents an ARI score greater than~0.5. The nodes are 
positioned by simulating a force-directed representation, based on the 
Fruchterman-Reingold force-directed algorithm~\cite{hagberg2008exploring}.

These graphs help visualize how other families are related to the four most-distinct 
families in our dataset. For example, Hotbar is almost equally distinguishable from all
other families. On the other hand, Adload is distinguishable from all families except 
Vobfus and BHO. This means that Adload, BHO, and Vobfus are mostly similar to each 
other, but highly distinguishable when clustered with other families in the dataset. 

It is also interesting to note that there are many families our dataset with extremely
poor pairwise clustering results. We see that Agent, CeeInject, Cycbot, DelfInject,
FakeRean, Obfuscator, Rbot, Renos, Vundo, Winwebsec, and Zbot all have
average ARI scores below~0.23. This indicates that there is a large subset
of the families that are virtually indistinguishable from each other.

\subsection{Clustering Families of Different Type}

In this set of clustering experiments, we consider seven families, each of which is of a different 
malware type. Specifically, the seven families considered, and their type, are the following:

\begin{description}
    \item[\bf Agent]---  Trojan
    \item[\bf Ceeinject]--- VirTool
    \item[\bf Cycbot]--- Backdoor
    \item[\bf FakeRean]--- Rogue
    \item[\bf Hotbar]--- Adware
    \item[\bf Lolyda]--- Password Stealer
    \item[\bf Vobfus]--- Worm
\end{description}

\begin{figure}[!htb]
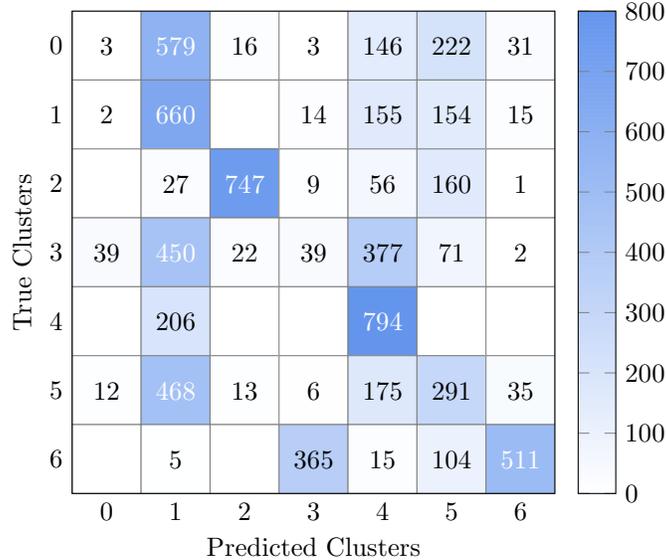

\centering
\input figures/conf7.tex
\caption{Clustering using seven families from different malware types}\label{fig:different9Fam}
\end{figure}

Figure~\ref{fig:different9Fam} shows the results of clustering these seven families, 
each of which belongs to a different malware type. We might expect well-defined clusters
in this case, but the ARI score is only~0.23, suggesting that a few families are still 
very similar, in spite of belonging to different malware types. 

The results in Figure~\ref{fig:different9Fam} indicates
that ``type'' is not a strong feature of malware. More specifically, 
we can say that the characteristics of bigrams that
distinguish one malware family from another are not strongly type-dependent.
This somewhat surprising result is useful, since it shows that
attempts to identify malware by generic type are, in general, unlikely to be successful,
at least when the analysis is based on byte bigram features.
However, we note in passing that the authors in~\cite{pirscoveanu2015analysis} use API 
calls as features and appear to have successfully classified selected malware by 
its type. Hence, it may be possible to obtain better results for malware type
by using other features.

\subsection{Clustering Families of the Same Type}

\begin{figure}[!htb]
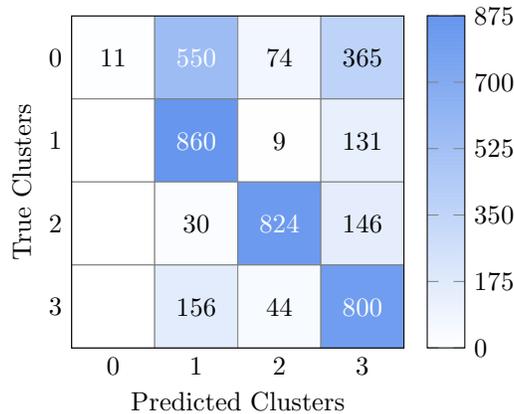

\centering
\input figures/conf4.tex
\caption{Clustering four Trojan families (Agent, Alureon, BHO, and Startpage)}\label{fig:sameFam1}
\end{figure}

In this section, we conducted two experiments to examine how well $K$-means 
clustering can distinguish between families belonging to the same malware type. 
Figure~\ref{fig:sameFam1} illustrates the results of clustering four families, all of which 
are Trojans---the specific families considered in this case are Agent, Alureon, BHO, and Startpage.
This result suggests that there are three well-defined clusters among these four families. 
We obtain an ARI of~0.35
in this case, which, interestingly, is much higher than the result obtained for malware
samples of different types in the previous section.

\begin{figure}[!htb]
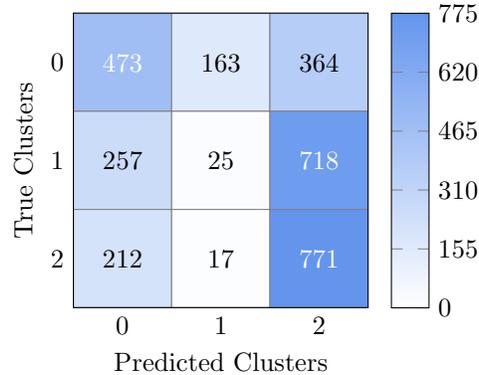

\centering
\input figures/conf3.tex
\caption{Clustering VirTool families (DelfInject, CeeInject, and Obfuscator)}\label{fig:sameFam2}
\end{figure}

Next, we cluster the three VirTool families in our dataset, namely, 
CeeInject, DelfInject, and Obfuscator.
In this, we obtain the results in Figure~\ref{fig:sameFam2}, which give us an ARI score 
of just~0.07. This number suggests a random clustering result, and implies that these
VirTool families are virtually indistinguishable, based on byte bigram features.

The results in this section indicate that the Trojan type is generic, in the sense that 
Trojan families can (and generally do) differ significantly from each other.
This is not surprising, as Trojan code tends to be dominated by the 
non-malicious part of the application, which would be expected to vary widely
between different Trojan families.

On the other hand, the VirTool type is highly specific, which results in an inability
to distinguish between these families. That is, the VirTool type is relatively homogeneous, 
making such samples difficult to distinguish from each other, even when they
are from different families.

\section{Conclusion and Future Work}\label{sec:con}

The goal of this research was to analyze malware clustering, with respect to 
families and types, based on elementary features and clustering techniques. 
We considered three different sets of experiments. 
In our first set of experiments, we clustered all families in pairs. In our second set
of experiments, we clustered seven families, with one family from each of the 
distinct types in our dataset. Finally, we conducted experiments where the 
clustered families all belong to the same malware type. All of our experiments
were based on $K$-means clustering using byte bigram features.

Our findings indicate that the relationship between malware families and malware type is
somewhat complex. This is not entirely unexpected, since some malware types, such as Trojans, 
are only very loosely related, while other types, such as VirTool, are much more specific. 
Indeed, we did find that families of the Trojan type were far easier to distinguish
from each other based on clustering, as compared to VirTool families.

More generally, our pairwise clustering results---and, in particular, the heatmap of ARI scores
generated from these pairwise clusters---enabled us to draw many conclusions concerning
similarities and differences between families. We were able to clearly see which families
were most distinct from all other families, and which subsets of families were the most
similar to each other. These results show that elementary cluster analysis is extremely useful 
for exploring relationships between malware families, and that such analysis could serve as
a guide for additional (and more costly) analysis of a given malware dataset.

In this paper, we performed cluster analysis to examine the relationship between malware families. 
Our focus was clustering using $K$-means and byte bigram features. For future work,
it would be interesting to consider larger numbers of clusters, and
to explore other clustering techniques, including
Gaussian mixture models, hierarchical techniques, 
spectral clustering, and density-based clustering. While $K$-means can be viewed as generating
``circular'' or ``spherical'' clusters, other techniques can produce clusters of more general shapes.
In addition, it would be interesting to experiment with other features, such as opcodes and API
call sequences. 


\bibliographystyle{plain}

\bibliography{references.bib,referencesMulti.bib}

\end{document}

%% file: preamble.tex
\usepackage{amsmath, amsfonts, amssymb, amsxtra, amsopn}
\usepackage{pgfplots}
\pgfplotsset{compat=1.13}
\usepackage{pgfplotstable}
\usepackage{graphicx}
\usepackage{multirow}
\usepackage{algorithm}
\usepackage{algorithmicx}
\usepackage{algpseudocode}
\usepackage{booktabs}
\usepackage{listings}
\usepackage{cmap}
\usepackage{colortbl}
\usepackage{adjustbox}
\usepackage{epsfig}

\usepackage[tableposition=top]{caption}

\PassOptionsToPackage{hyphens}{url}

\usepackage{hyperref}
\hypersetup{colorlinks=true,linkcolor=black,citecolor=black,urlcolor=blue,filecolor=black}
\hypersetup{pdfpagemode=UseNone,pdfstartview=}

\pgfkeys{
    /pgf/number format/precision=4, 
    /pgf/number format/fixed zerofill=true }
    
\pgfplotstableset{
    /color cells/min/.initial=0,
    /color cells/max/.initial=1000,
    /color cells/textcolor/.initial=,
    color cells/.code={%
        \pgfqkeys{/color cells}{#1}%
        \pgfkeysalso{%
            postproc cell content/.code={%
                \begingroup
                \pgfkeysgetvalue{/pgfplots/table/@preprocessed cell content}\value
\ifx\value\empty
\endgroup
\else
                \pgfmathfloatparsenumber{\value}%
                \pgfmathfloattofixed{\pgfmathresult}%
                \let\value=\pgfmathresult
                \pgfplotscolormapaccess
                    [\pgfkeysvalueof{/color cells/min}:\pgfkeysvalueof{/color cells/max}]%
                    {\value}%
                    {\pgfkeysvalueof{/pgfplots/colormap name}}%
                \pgfkeysgetvalue{/pgfplots/table/@cell content}\typesetvalue
                \pgfkeysgetvalue{/color cells/textcolor}\textcolorvalue
                \toks0=\expandafter{\typesetvalue}%
                \xdef\temp{%
                    \noexpand\pgfkeysalso{%
                        @cell content={%
                            \noexpand\cellcolor[rgb]{\pgfmathresult}%
                            \noexpand\definecolor{mapped color}{rgb}{\pgfmathresult}%
                            \ifx\textcolorvalue\empty
                            \else
                                \noexpand\color{\textcolorvalue}%
                            \fi
                            \the\toks0 %
                        }%
                    }%
                }%
                \endgroup
                \temp
\fi
            }%
        }%
    }
}

\usepackage{enumitem}
\setlist[itemize]{noitemsep, topsep=0pt}

\advance\oddsidemargin by -0.2in
\advance\textwidth by 0.4in

\advance\topmargin by -0.25in
\advance\textheight by 0.5in

\def\eref#1{(\ref{#1})}

\long\def\symbolfootnotetext[#1]#2{\begingroup%
\def\thefootnote{\fnsymbol{footnote}}\footnotetext[#1]{#2}\endgroup}

\DeclareMathOperator{\RI}{RI}

\def\centroid#1{\mbox{centroid}(#1)}
\def\distortion#1{\mbox{distortion}_{#1}}

\algrenewcommand{\algorithmiccomment}[1]{{\color{red}{\tt //}\ #1}}

\algnewcommand{\Initialize}[1]{%
  \State \textbf{Initialize:}
  \Statex \hspace*{\algorithmicindent}\parbox[t]{.8\linewidth}{\raggedright #1}
}
\algnewcommand{\Given}[1]{%
  \State \textbf{Given:}
  \Statex \hspace*{\algorithmicindent}\parbox[t]{.8\linewidth}{\raggedright #1}
}

%% file: figures/figure_3.tex
    \begin{tikzpicture}[scale=0.8]
    %
    %
    \draw[thick,color=black,fill=black] (0.8,2.7) circle (0.08);
    \draw[thick,color=black,fill=black] (1.0,3.0) circle (0.08);
    \draw[thick,color=black,fill=black] (1.2,2.5) circle (0.08);
    \draw[thick,color=black,fill=black] (1.35,3.35) circle (0.08);
    \draw[thick,color=black,fill=black] (0.75,3.45) circle (0.08);
    \draw[thick,color=black,fill=black] (1.75,4.0) circle (0.08);
    \draw[thick,color=black,fill=black] (0.5,3.6) circle (0.08);
    \draw[thick,color=black,fill=black] (1.0,4.1) circle (0.08);
    
    \draw[thick,color=black,fill=black] (2.0,1.35) circle (0.08);
    \draw[thick,color=black,fill=black] (1.5,0.8) circle (0.08);
    \draw[thick,color=black,fill=black] (1.15,1.65) circle (0.08);
    \draw[thick,color=black,fill=black] (2.25,1.15) circle (0.08);
    \draw[thick,color=black,fill=black] (2.05,0.5) circle (0.08);

    \draw[thick,color=black,fill=black] (4.0,2.0) circle (0.08);
    \draw[thick,color=black,fill=black] (4.25,2.1) circle (0.08);
    \draw[thick,color=black,fill=black] (4.6,2.3) circle (0.08);
    \draw[thick,color=black,fill=black] (5.25,3.0) circle (0.08);
    \draw[thick,color=black,fill=black] (4.75,3.0) circle (0.08);
    \draw[thick,color=black,fill=black] (4.0,2.75) circle (0.08);
    \draw[thick,color=black,fill=black] (3.75,3.05) circle (0.08);
    \draw[thick,color=black,fill=black] (4.6,3.35) circle (0.08);
    \draw[thick,color=black,fill=black] (5.0,3.2) circle (0.08);
    
    \draw[blue,thick] plot [smooth] coordinates {
    (0.75,1.5) (1.0,0.5) (2.0,0.25) (2.75,1.0) (2.25,1.75) (1,1.9) (0.75,1.75) (0.75,1.5)};
    \draw[red,thick] plot [smooth] coordinates {
    (4.0,1.65) (5.5,2.5) (5.6,3.5) (4.75,3.75) (3.55,3.25) (3.5,2.25) (3.75,1.7) (4.0,1.65)};
    \draw[green,thick] plot [smooth] coordinates {
    (0.5,2.375) (1.5,2.25) (2.15,3.5) (2.0,4.22) (1.0,4.5) (0.25,3.5) (0.5,2.375)};
        
        
     \draw[thick,color=black,->] (0,0) -- (6,0); 
     \draw[thick,color=black,->] (0,0) -- (0,5); 

    %
    %
    \draw[thick,color=black,fill=black] (9.8,2.7) circle (0.08);
    \draw[thick,color=black,fill=black] (10.0,3.0) circle (0.08);
    \draw[thick,color=black,fill=black] (10.2,2.5) circle (0.08);
    \draw[thick,color=black,fill=black] (10.35,3.35) circle (0.08);
    \draw[thick,color=black,fill=black] (9.75,3.45) circle (0.08);
    \draw[thick,color=black,fill=black] (9.5,3.6) circle (0.08);
    
    \draw[thick,color=black,fill=black] (11.0,1.35) circle (0.08);
    \draw[thick,color=black,fill=black] (10.5,0.8) circle (0.08);
    \draw[thick,color=black,fill=black](10.15,1.65) circle (0.08);
    \draw[thick,color=black,fill=black] (11.25,1.15) circle (0.08);
    \draw[thick,color=black,fill=black] (11.05,0.5) circle (0.08);

    \draw[thick,color=black,fill=black] (13.0,2.0) circle (0.08);
    \draw[thick,color=black,fill=black] (13.25,2.1) circle (0.08);
    \draw[thick,color=black,fill=black] (13.6,2.3) circle (0.08);
    \draw[thick,color=black,fill=black] (14.25,3.0) circle (0.08);
    \draw[thick,color=black,fill=black] (14.25,3.0) circle (0.08);
    \draw[thick,color=black,fill=black] (13.75,3.0) circle (0.08);
    \draw[thick,color=black,fill=black] (13.0,2.75) circle (0.08);
    \draw[thick,color=black,fill=black] (12.75,3.05) circle (0.08);
    \draw[thick,color=black,fill=black] (13.6,3.35) circle (0.08);
    \draw[thick,color=black,fill=black] (14.0,3.2) circle (0.08);

    \draw[thick,color=black,fill=black] (10.75,4.0) circle (0.08);
    \draw[thick,color=black,fill=black] (10.0,4.1) circle (0.08);

    \draw[blue,thick] plot [smooth] coordinates {
    (9.75,1.5) (10.0,0.5) (11.0,0.25) (11.75,1.0) (11.25,1.75) (10,1.9) (9.75,1.75) (9.75,1.5)};
    \draw[red,thick] plot [smooth] coordinates {
    (13.0,1.65) (14.5,2.5) (14.6,3.5) (13.75,3.75) (11.0,4.22) (10.0,4.5)  (9.85,3.9) (11,3.625)
    (12.5,2.2) (12.75,1.8) (13.0,1.65)}; 
    \draw[green,thick] plot [smooth] coordinates {
    (9.5,2.375) (10.5,2.25) (11.0,3.0) (10.75,3.5) (10.0,3.75) (9.5,3.9) (9.25,3.25) (9.5,2.375)};


     \draw[thick,color=black,->] (9,0) -- (15,0); 
     \draw[thick,color=black,->] (9,0) -- (9,5); 
   
    \end{tikzpicture}

%% file: figures/elbow_d.tex
\begin{tikzpicture}[scale=0.55]
\begin{axis}[no markers,smooth,
		   width=0.75\textwidth,
		   height=0.6\textwidth,
	 	   x tick label style={
   		 	/pgf/number format/.cd,
			/pgf/number format/1000 sep={},
   			fixed,
   			fixed zerofill,
    			precision=0
		   },
	 	   y tick label style={
    		 	/pgf/number format/.cd,
   			fixed,
   			fixed zerofill,
    			precision=3
		    },
                    xmin=1,xmax=20,
                    ymin=0.045,ymax=0.23,
                    legend pos=north east,
                    legend cell align={left},
                    xtick={2,4,6,8,10,12,14,16,18,20},
                    ytick={0.050,0.075,0.100,0.125,0.150,0.175,0.200,0.225},
                    xlabel={$K$},
                    ylabel={Distortion}] 
\addplot[color=red,thick] coordinates {
(1,0.22838167726841788)
(2,0.13414876903195008)
(3,0.11779415000192103)
(4,0.09807172435963532)
(5,0.08646766041855718)
(6,0.0796018459535686)
(7,0.07911008651958301)
(8,0.0731730403640641)
(9,0.07210437447002757)
(10,0.06733560412182697)
(11,0.06683278928041589)
(12,0.06400373286867632)
(13,0.06066698440825672)
(14,0.061902577565012704)
(15,0.05818017044891091)
(16,0.056817918458590314)
(17,0.05554729843902755)
(18,0.05363162033854852)
(19,0.05163990470650576)
(20,0.050854234978014855)
};
\end{axis}
\end{tikzpicture}

%% file: figures/elbow_i.tex
\begin{tikzpicture}[scale=0.55]
\begin{axis}[no markers,smooth,
		   width=0.75\textwidth,
		   height=0.6\textwidth,
	 	   x tick label style={
   		 	/pgf/number format/.cd,
			/pgf/number format/1000 sep={},
   			fixed,
   			fixed zerofill,
    			precision=0
		   },
	 	   y tick label style={
    		 	/pgf/number format/.cd,
   			fixed,
   			fixed zerofill,
			1000 sep={},
    			precision=0
		    },
                    xmin=1,xmax=20,
                    ymin=58,ymax=1400,
                    legend pos=north east,
                    legend cell align={left},
                    xtick={2,4,6,8,10,12,14,16,18,20},
                    ytick={100,250,400,550,700,850,1000,1150,1300},
                    xlabel={$K$},
                    ylabel={Inertia}] 
\addplot[color=red,thick] coordinates {
(1,1380.1437070000698)
(2,551.0376811475447)
(3,378.4657616496158)
(4,281.6858568611039)
(5,236.4959367545155)
(6,203.36974434842097)
(7,186.5249636625105)
(8,171.4832193781069)
(9,159.4322482728705)
(10,148.11641094816108)
(11,136.87433372573318)
(12,125.89404277581349)
(13,113.81058075365617)
(14,111.13673347826744)
(15,101.71351212676115)
(16,95.89645236401158)
(17,92.02115600709718)
(18,85.76380770322734)
(19,82.0170066105975)
(20,78.18765591524217)
};
\end{axis}
\end{tikzpicture}

%% file: figures/conf2a.tex
\begin{tikzpicture}[scale=1.0]
    \begin{axis}[
    	title={\scriptsize$\mbox{ARI} = 0.1777$},
	title style={at={(0.5,0.875)}},
        width=3.0cm,
        height=3.0cm,
    colormap={bluewhite}{color=(white) rgb255=(100,149,237)},
        xticklabels={0,1},
        xtick={0,...,1},
        xtick style={draw=none},
        yticklabels={0,1},
        ytick={0,...,1},
        ytick style={draw=none},
        enlargelimits=false,
        xlabel style={font=\scriptsize},
        ylabel style={font=\scriptsize},
        xticklabel style={font=\scriptsize},
        yticklabel style={font=\scriptsize},
        point meta min=0,
        point meta max=850,
        nodes near coords={\pgfmathprintnumber\pgfplotspointmeta},
        nodes near coords black white/.style={
            small value/.style={
                font=\scriptsize,
                yshift=-7pt,
                text=black,
                /pgf/number format/fixed,
                /pgf/number format/precision=0,
                /pgf/number format/zerofill
            },
            large value/.style={
                font=\scriptsize,
                yshift=-7pt,
                text=white,
                /pgf/number format/fixed,
                /pgf/number format/precision=0,
                /pgf/number format/zerofill
            },
            every node near coord/.style={
                check for zero/.code={
                    \pgfmathfloatifflags{\pgfplotspointmeta}{0}{
                        \pgfkeys{/tikz/coordinate}
                    }{
                        \begingroup
                        \pgfkeys{/pgf/fpu}
                        \pgfmathparse{\pgfplotspointmeta<#1}
                        \global\let\result=\pgfmathresult
                        \endgroup
                        %
                        %
                        \pgfmathfloatcreate{1}{1.0}{0}
                        \let\ONE=\pgfmathresult
                        \ifx\result\ONE
                            \pgfkeysalso{/pgfplots/small value}
                        \else
                            \pgfkeysalso{/pgfplots/large value}
                        \fi
                    }
                },
                check for zero,
            },
        },
        nodes near coords black white=420,
    ]
        \addplot[
            matrix plot,
            mesh/cols=2,
            point meta=explicit,draw=gray
        ] table [meta=C] {
            x y C
0 0 587
1 0 413
0 1 165
1 1 835
};
    \end{axis}
\end{tikzpicture}
%

%% file: figures/conf2b.tex
\begin{tikzpicture}[scale=1.0]
    \begin{axis}[
    	title={\scriptsize$\mbox{ARI} = 0.1932$},
	title style={at={(0.5,0.875)}},
        width=3.0cm,
        height=3.0cm,
    colormap={bluewhite}{color=(white) rgb255=(100,149,237)},
        xticklabels={0,1},
        xtick={0,...,1},
        xtick style={draw=none},
        yticklabels={0,1},
        ytick={0,...,1},
        ytick style={draw=none},
        enlargelimits=false,
        xlabel style={font=\scriptsize},
        ylabel style={font=\scriptsize},
        xticklabel style={font=\scriptsize},
        yticklabel style={font=\scriptsize},
        point meta min=0,
        point meta max=850,
        nodes near coords={\pgfmathprintnumber\pgfplotspointmeta},
        nodes near coords black white/.style={
            small value/.style={
                font=\scriptsize,
                yshift=-7pt,
                text=black,
                /pgf/number format/fixed,
                /pgf/number format/precision=0,
                /pgf/number format/zerofill
            },
            large value/.style={
                font=\scriptsize,
                yshift=-7pt,
                text=white,
                /pgf/number format/fixed,
                /pgf/number format/precision=0,
                /pgf/number format/zerofill
            },
            every node near coord/.style={
                check for zero/.code={
                    \pgfmathfloatifflags{\pgfplotspointmeta}{0}{
                        \pgfkeys{/tikz/coordinate}
                    }{
                        \begingroup
                        \pgfkeys{/pgf/fpu}
                        \pgfmathparse{\pgfplotspointmeta<#1}
                        \global\let\result=\pgfmathresult
                        \endgroup
                        %
                        %
                        \pgfmathfloatcreate{1}{1.0}{0}
                        \let\ONE=\pgfmathresult
                        \ifx\result\ONE
                            \pgfkeysalso{/pgfplots/small value}
                        \else
                            \pgfkeysalso{/pgfplots/large value}
                        \fi
                    }
                },
                check for zero,
            },
        },
        nodes near coords black white=420,
    ]
        \addplot[
            matrix plot,
            mesh/cols=2,
            point meta=explicit,draw=gray
        ] table [meta=C] {
            x y C
0 0 900
1 0 100
0 1 460
1 1 540
};
    \end{axis}
\end{tikzpicture}
%

%% file: figures/conf2c.tex
\begin{tikzpicture}[scale=1.0]
    \begin{axis}[
    	title={\scriptsize$\mbox{ARI} = 0.1969$},
	title style={at={(0.5,0.875)}},
        width=3.0cm,
        height=3.0cm,
    colormap={bluewhite}{color=(white) rgb255=(100,149,237)},
        xticklabels={0,1},
        xtick={0,...,1},
        xtick style={draw=none},
        yticklabels={0,1},
        ytick={0,...,1},
        ytick style={draw=none},
        enlargelimits=false,
        xlabel style={font=\scriptsize},
        ylabel style={font=\scriptsize},
        xticklabel style={font=\scriptsize},
        yticklabel style={font=\scriptsize},
        point meta min=0,
        point meta max=850,
        nodes near coords={\pgfmathprintnumber\pgfplotspointmeta},
        nodes near coords black white/.style={
            small value/.style={
                font=\scriptsize,
                yshift=-7pt,
                text=black,
                /pgf/number format/fixed,
                /pgf/number format/precision=0,
                /pgf/number format/zerofill,
                /pgf/number format/1000 sep={}
            },
            large value/.style={
                font=\scriptsize,
                yshift=-7pt,
                text=white,
                /pgf/number format/fixed,
                /pgf/number format/precision=0,
                /pgf/number format/zerofill,
                /pgf/number format/1000 sep={}
            },
            every node near coord/.style={
                check for zero/.code={
                    \pgfmathfloatifflags{\pgfplotspointmeta}{0}{
                        \pgfkeys{/tikz/coordinate}
                    }{
                        \begingroup
                        \pgfkeys{/pgf/fpu}
                        \pgfmathparse{\pgfplotspointmeta<#1}
                        \global\let\result=\pgfmathresult
                        \endgroup
                        %
                        %
                        \pgfmathfloatcreate{1}{1.0}{0}
                        \let\ONE=\pgfmathresult
                        \ifx\result\ONE
                            \pgfkeysalso{/pgfplots/small value}
                        \else
                            \pgfkeysalso{/pgfplots/large value}
                        \fi
                    }
                },
                check for zero,
            },
        },
        nodes near coords black white=420,
    ]
        \addplot[
            matrix plot,
            mesh/cols=2,
            point meta=explicit,draw=gray
        ] table [meta=C] {
            x y C
0 0 1000
1 0 0
0 1 556
1 1 444
};
    \end{axis}
\end{tikzpicture}
%

%% file: figures/conf2d.tex
\begin{tikzpicture}[scale=1.0]
    \begin{axis}[
    	title={\scriptsize$\mbox{ARI} = 0.2048$},
	title style={at={(0.5,0.875)}},
        width=3.0cm,
        height=3.0cm,
    colormap={bluewhite}{color=(white) rgb255=(100,149,237)},
        xticklabels={0,1},
        xtick={0,...,1},
        xtick style={draw=none},
        yticklabels={0,1},
        ytick={0,...,1},
        ytick style={draw=none},
        enlargelimits=false,
        xlabel style={font=\scriptsize},
        ylabel style={font=\scriptsize},
        xticklabel style={font=\scriptsize},
        yticklabel style={font=\scriptsize},
        point meta min=0,
        point meta max=850,
        nodes near coords={\pgfmathprintnumber\pgfplotspointmeta},
        nodes near coords black white/.style={
            small value/.style={
                font=\scriptsize,
                yshift=-7pt,
                text=black,
                /pgf/number format/fixed,
                /pgf/number format/precision=0,
                /pgf/number format/zerofill
            },
            large value/.style={
                font=\scriptsize,
                yshift=-7pt,
                text=white,
                /pgf/number format/fixed,
                /pgf/number format/precision=0,
                /pgf/number format/zerofill
            },
            every node near coord/.style={
                check for zero/.code={
                    \pgfmathfloatifflags{\pgfplotspointmeta}{0}{
                        \pgfkeys{/tikz/coordinate}
                    }{
                        \begingroup
                        \pgfkeys{/pgf/fpu}
                        \pgfmathparse{\pgfplotspointmeta<#1}
                        \global\let\result=\pgfmathresult
                        \endgroup
                        %
                        %
                        \pgfmathfloatcreate{1}{1.0}{0}
                        \let\ONE=\pgfmathresult
                        \ifx\result\ONE
                            \pgfkeysalso{/pgfplots/small value}
                        \else
                            \pgfkeysalso{/pgfplots/large value}
                        \fi
                    }
                },
                check for zero,
            },
        },
        nodes near coords black white=420,
    ]
        \addplot[
            matrix plot,
            mesh/cols=2,
            point meta=explicit,draw=gray
        ] table [meta=C] {
            x y C
0 0 831
1 0 169
0 1 378
1 1 622
};
    \end{axis}
\end{tikzpicture}
%

%% file: figures/conf2e_cb.tex
\begin{tikzpicture}[scale=1.0]
    \begin{axis}[
    	title={\scriptsize$\mbox{ARI} = 0.2168$},
	title style={at={(0.5,0.875)}},
        width=3.0cm,
        height=3.0cm,
    colormap={bluewhite}{color=(white) rgb255=(100,149,237)},
        xticklabels={0,1},
        xtick={0,...,1},
        xtick style={draw=none},
        yticklabels={0,1},
        ytick={0,...,1},
        ytick style={draw=none},
        enlargelimits=false,
        xlabel style={font=\scriptsize},
        ylabel style={font=\scriptsize},
        xticklabel style={font=\scriptsize},
        yticklabel style={font=\scriptsize},
        colorbar,
        colorbar style={
            ytick={0,250,500,750,1000},
            yticklabels={0,250,500,750,1000},
            at={(1.1,1.0)},
            width=5pt,
            yticklabel={\pgfmathprintnumber\tick},
            yticklabel style={font=\scriptsize,
                    /pgf/number format/fixed,
            /pgf/number format/precision=0,
            /pgf/number format/1000 sep={}}
        },
        point meta min=0,
        point meta max=1000,
        nodes near coords={\pgfmathprintnumber\pgfplotspointmeta},
        nodes near coords black white/.style={
            small value/.style={
                font=\scriptsize,
                yshift=-7pt,
                text=black,
                /pgf/number format/fixed,
                /pgf/number format/precision=0,
                /pgf/number format/zerofill
            },
            large value/.style={
                font=\scriptsize,
                yshift=-7pt,
                text=white,
                /pgf/number format/fixed,
                /pgf/number format/precision=0,
                /pgf/number format/zerofill
            },
            every node near coord/.style={
                check for zero/.code={
                    \pgfmathfloatifflags{\pgfplotspointmeta}{0}{
                        \pgfkeys{/tikz/coordinate}
                    }{
                        \begingroup
                        \pgfkeys{/pgf/fpu}
                        \pgfmathparse{\pgfplotspointmeta<#1}
                        \global\let\result=\pgfmathresult
                        \endgroup
                        %
                        %
                        \pgfmathfloatcreate{1}{1.0}{0}
                        \let\ONE=\pgfmathresult
                        \ifx\result\ONE
                            \pgfkeysalso{/pgfplots/small value}
                        \else
                            \pgfkeysalso{/pgfplots/large value}
                        \fi
                    }
                },
                check for zero,
            },
        },
        nodes near coords black white=420,
    ]
        \addplot[
            matrix plot,
            mesh/cols=2,
            point meta=explicit,draw=gray
        ] table [meta=C] {
            x y C
0 0 701
1 0 299
0 1 235
1 1 765
};
    \end{axis}
\end{tikzpicture}
%

%% file: figures/conf2f.tex
\begin{tikzpicture}[scale=1.0]
    \begin{axis}[
    	title={\scriptsize$\mbox{ARI} = 0.2272$},
	title style={at={(0.5,0.875)}},
        width=3.0cm,
        height=3.0cm,
    colormap={bluewhite}{color=(white) rgb255=(100,149,237)},
        xticklabels={0,1},
        xtick={0,...,1},
        xtick style={draw=none},
        yticklabels={0,1},
        ytick={0,...,1},
        ytick style={draw=none},
        enlargelimits=false,
        xlabel style={font=\scriptsize},
        ylabel style={font=\scriptsize},
        xticklabel style={font=\scriptsize},
        yticklabel style={font=\scriptsize},
        point meta min=0,
        point meta max=850,
        nodes near coords={\pgfmathprintnumber\pgfplotspointmeta},
        nodes near coords black white/.style={
            small value/.style={
                font=\scriptsize,
                yshift=-7pt,
                text=black,
                /pgf/number format/fixed,
                /pgf/number format/precision=0,
                /pgf/number format/zerofill
            },
            large value/.style={
                font=\scriptsize,
                yshift=-7pt,
                text=white,
                /pgf/number format/fixed,
                /pgf/number format/precision=0,
                /pgf/number format/zerofill
            },
            every node near coord/.style={
                check for zero/.code={
                    \pgfmathfloatifflags{\pgfplotspointmeta}{0}{
                        \pgfkeys{/tikz/coordinate}
                    }{
                        \begingroup
                        \pgfkeys{/pgf/fpu}
                        \pgfmathparse{\pgfplotspointmeta<#1}
                        \global\let\result=\pgfmathresult
                        \endgroup
                        %
                        %
                        \pgfmathfloatcreate{1}{1.0}{0}
                        \let\ONE=\pgfmathresult
                        \ifx\result\ONE
                            \pgfkeysalso{/pgfplots/small value}
                        \else
                            \pgfkeysalso{/pgfplots/large value}
                        \fi
                    }
                },
                check for zero,
            },
        },
        nodes near coords black white=420,
    ]
        \addplot[
            matrix plot,
            mesh/cols=2,
            point meta=explicit,draw=gray
        ] table [meta=C] {
            x y C
0 0 631
1 0 369
0 1 154
1 1 846
};
    \end{axis}
\end{tikzpicture}
%

%% file: figures/conf2g.tex
\begin{tikzpicture}[scale=1.0]
    \begin{axis}[
    	title={\scriptsize$\mbox{ARI} = 0.2467$},
	title style={at={(0.5,0.875)}},
        width=3.0cm,
        height=3.0cm,
    colormap={bluewhite}{color=(white) rgb255=(100,149,237)},
        xticklabels={0,1},
        xtick={0,...,1},
        xtick style={draw=none},
        yticklabels={0,1},
        ytick={0,...,1},
        ytick style={draw=none},
        enlargelimits=false,
        xlabel style={font=\scriptsize},
        ylabel style={font=\scriptsize},
        xticklabel style={font=\scriptsize},
        yticklabel style={font=\scriptsize},
        point meta min=0,
        point meta max=850,
        nodes near coords={\pgfmathprintnumber\pgfplotspointmeta},
        nodes near coords black white/.style={
            small value/.style={
                font=\scriptsize,
                yshift=-7pt,
                text=black,
                /pgf/number format/fixed,
                /pgf/number format/precision=0,
                /pgf/number format/zerofill
            },
            large value/.style={
                font=\scriptsize,
                yshift=-7pt,
                text=white,
                /pgf/number format/fixed,
                /pgf/number format/precision=0,
                /pgf/number format/zerofill
            },
            every node near coord/.style={
                check for zero/.code={
                    \pgfmathfloatifflags{\pgfplotspointmeta}{0}{
                        \pgfkeys{/tikz/coordinate}
                    }{
                        \begingroup
                        \pgfkeys{/pgf/fpu}
                        \pgfmathparse{\pgfplotspointmeta<#1}
                        \global\let\result=\pgfmathresult
                        \endgroup
                        %
                        %
                        \pgfmathfloatcreate{1}{1.0}{0}
                        \let\ONE=\pgfmathresult
                        \ifx\result\ONE
                            \pgfkeysalso{/pgfplots/small value}
                        \else
                            \pgfkeysalso{/pgfplots/large value}
                        \fi
                    }
                },
                check for zero,
            },
        },
        nodes near coords black white=420,
    ]
        \addplot[
            matrix plot,
            mesh/cols=2,
            point meta=explicit,draw=gray
        ] table [meta=C] {
            x y C
0 0 951
1 0 49
0 1 454
1 1 546
};
    \end{axis}
\end{tikzpicture}
%

%% file: figures/conf2h.tex
\begin{tikzpicture}[scale=1.0]
    \begin{axis}[
    	title={\scriptsize$\mbox{ARI} = 0.2476$},
	title style={at={(0.5,0.875)}},
        width=3.0cm,
        height=3.0cm,
    colormap={bluewhite}{color=(white) rgb255=(100,149,237)},
        xticklabels={0,1},
        xtick={0,...,1},
        xtick style={draw=none},
        yticklabels={0,1},
        ytick={0,...,1},
        ytick style={draw=none},
        enlargelimits=false,
        xlabel style={font=\scriptsize},
        ylabel style={font=\scriptsize},
        xticklabel style={font=\scriptsize},
        yticklabel style={font=\scriptsize},
        point meta min=0,
        point meta max=850,
        nodes near coords={\pgfmathprintnumber\pgfplotspointmeta},
        nodes near coords black white/.style={
            small value/.style={
                font=\scriptsize,
                yshift=-7pt,
                text=black,
                /pgf/number format/fixed,
                /pgf/number format/precision=0,
                /pgf/number format/zerofill
            },
            large value/.style={
                font=\scriptsize,
                yshift=-7pt,
                text=white,
                /pgf/number format/fixed,
                /pgf/number format/precision=0,
                /pgf/number format/zerofill
            },
            every node near coord/.style={
                check for zero/.code={
                    \pgfmathfloatifflags{\pgfplotspointmeta}{0}{
                        \pgfkeys{/tikz/coordinate}
                    }{
                        \begingroup
                        \pgfkeys{/pgf/fpu}
                        \pgfmathparse{\pgfplotspointmeta<#1}
                        \global\let\result=\pgfmathresult
                        \endgroup
                        %
                        %
                        \pgfmathfloatcreate{1}{1.0}{0}
                        \let\ONE=\pgfmathresult
                        \ifx\result\ONE
                            \pgfkeysalso{/pgfplots/small value}
                        \else
                            \pgfkeysalso{/pgfplots/large value}
                        \fi
                    }
                },
                check for zero,
            },
        },
        nodes near coords black white=420,
    ]
        \addplot[
            matrix plot,
            mesh/cols=2,
            point meta=explicit,draw=gray
        ] table [meta=C] {
            x y C
0 0 861
1 0 139
0 1 363
1 1 637
};
    \end{axis}
\end{tikzpicture}
%

%% file: figures/conf2i.tex
\begin{tikzpicture}[scale=1.0]
    \begin{axis}[
    	title={\scriptsize$\mbox{ARI} = 0.2588$},
	title style={at={(0.5,0.875)}},
        width=3.0cm,
        height=3.0cm,
    colormap={bluewhite}{color=(white) rgb255=(100,149,237)},
        xticklabels={0,1},
        xtick={0,...,1},
        xtick style={draw=none},
        yticklabels={0,1},
        ytick={0,...,1},
        ytick style={draw=none},
        enlargelimits=false,
        xlabel style={font=\scriptsize},
        ylabel style={font=\scriptsize},
        xticklabel style={font=\scriptsize},
        yticklabel style={font=\scriptsize},
        point meta min=0,
        point meta max=850,
        nodes near coords={\pgfmathprintnumber\pgfplotspointmeta},
        nodes near coords black white/.style={
            small value/.style={
                font=\scriptsize,
                yshift=-7pt,
                text=black,
                /pgf/number format/fixed,
                /pgf/number format/precision=0,
                /pgf/number format/zerofill
            },
            large value/.style={
                font=\scriptsize,
                yshift=-7pt,
                text=white,
                /pgf/number format/fixed,
                /pgf/number format/precision=0,
                /pgf/number format/zerofill
            },
            every node near coord/.style={
                check for zero/.code={
                    \pgfmathfloatifflags{\pgfplotspointmeta}{0}{
                        \pgfkeys{/tikz/coordinate}
                    }{
                        \begingroup
                        \pgfkeys{/pgf/fpu}
                        \pgfmathparse{\pgfplotspointmeta<#1}
                        \global\let\result=\pgfmathresult
                        \endgroup
                        %
                        %
                        \pgfmathfloatcreate{1}{1.0}{0}
                        \let\ONE=\pgfmathresult
                        \ifx\result\ONE
                            \pgfkeysalso{/pgfplots/small value}
                        \else
                            \pgfkeysalso{/pgfplots/large value}
                        \fi
                    }
                },
                check for zero,
            },
        },
        nodes near coords black white=420,
    ]
        \addplot[
            matrix plot,
            mesh/cols=2,
            point meta=explicit,draw=gray
        ] table [meta=C] {
            x y C
0 0 951
1 0 49
0 1 442
1 1 558
};
    \end{axis}
\end{tikzpicture}
%

%% file: figures/conf2j_cb.tex
\begin{tikzpicture}[scale=1.0]
    \begin{axis}[
    	title={\scriptsize$\mbox{ARI} = 0.2669$},
	title style={at={(0.5,0.875)}},
        width=3.0cm,
        height=3.0cm,
    colormap={bluewhite}{color=(white) rgb255=(100,149,237)},
        xticklabels={0,1},
        xtick={0,...,1},
        xtick style={draw=none},
        yticklabels={0,1},
        ytick={0,...,1},
        ytick style={draw=none},
        enlargelimits=false,
        xlabel style={font=\scriptsize},
        ylabel style={font=\scriptsize},
        xticklabel style={font=\scriptsize},
        yticklabel style={font=\scriptsize},
        colorbar,
        colorbar style={
            ytick={0,250,500,750,1000},
            yticklabels={0,250,500,750,1000},
            at={(1.1,1.0)},
            width=5pt,
            yticklabel={\pgfmathprintnumber\tick},
            yticklabel style={font=\scriptsize,
                    /pgf/number format/fixed,
            /pgf/number format/precision=0,
            /pgf/number format/1000 sep={}}
        },
        point meta min=0,
        point meta max=1000,
        nodes near coords={\pgfmathprintnumber\pgfplotspointmeta},
        nodes near coords black white/.style={
            small value/.style={
                font=\scriptsize,
                yshift=-7pt,
                text=black,
                /pgf/number format/fixed,
                /pgf/number format/precision=0,
                /pgf/number format/zerofill
            },
            large value/.style={
                font=\scriptsize,
                yshift=-7pt,
                text=white,
                /pgf/number format/fixed,
                /pgf/number format/precision=0,
                /pgf/number format/zerofill
            },
            every node near coord/.style={
                check for zero/.code={
                    \pgfmathfloatifflags{\pgfplotspointmeta}{0}{
                        \pgfkeys{/tikz/coordinate}
                    }{
                        \begingroup
                        \pgfkeys{/pgf/fpu}
                        \pgfmathparse{\pgfplotspointmeta<#1}
                        \global\let\result=\pgfmathresult
                        \endgroup
                        %
                        %
                        \pgfmathfloatcreate{1}{1.0}{0}
                        \let\ONE=\pgfmathresult
                        \ifx\result\ONE
                            \pgfkeysalso{/pgfplots/small value}
                        \else
                            \pgfkeysalso{/pgfplots/large value}
                        \fi
                    }
                },
                check for zero,
            },
        },
        nodes near coords black white=420,
    ]
        \addplot[
            matrix plot,
            mesh/cols=2,
            point meta=explicit,draw=gray
        ] table [meta=C] {
            x y C
0 0 846
1 0 154
0 1 329
1 1 671
};
    \end{axis}
\end{tikzpicture}
%

%% file: figures/conf2k.tex
\begin{tikzpicture}[scale=1.0]
    \begin{axis}[
    	title={\scriptsize$\mbox{ARI} = 0.2711$},
	title style={at={(0.5,0.875)}},
        width=3.0cm,
        height=3.0cm,
    colormap={bluewhite}{color=(white) rgb255=(100,149,237)},
        xticklabels={0,1},
        xtick={0,...,1},
        xtick style={draw=none},
        yticklabels={0,1},
        ytick={0,...,1},
        ytick style={draw=none},
        enlargelimits=false,
        xlabel style={font=\scriptsize},
        ylabel style={font=\scriptsize},
        xticklabel style={font=\scriptsize},
        yticklabel style={font=\scriptsize},
        point meta min=0,
        point meta max=850,
        nodes near coords={\pgfmathprintnumber\pgfplotspointmeta},
        nodes near coords black white/.style={
            small value/.style={
                font=\scriptsize,
                yshift=-7pt,
                text=black,
                /pgf/number format/fixed,
                /pgf/number format/precision=0,
                /pgf/number format/zerofill
            },
            large value/.style={
                font=\scriptsize,
                yshift=-7pt,
                text=white,
                /pgf/number format/fixed,
                /pgf/number format/precision=0,
                /pgf/number format/zerofill
            },
            every node near coord/.style={
                check for zero/.code={
                    \pgfmathfloatifflags{\pgfplotspointmeta}{0}{
                        \pgfkeys{/tikz/coordinate}
                    }{
                        \begingroup
                        \pgfkeys{/pgf/fpu}
                        \pgfmathparse{\pgfplotspointmeta<#1}
                        \global\let\result=\pgfmathresult
                        \endgroup
                        %
                        %
                        \pgfmathfloatcreate{1}{1.0}{0}
                        \let\ONE=\pgfmathresult
                        \ifx\result\ONE
                            \pgfkeysalso{/pgfplots/small value}
                        \else
                            \pgfkeysalso{/pgfplots/large value}
                        \fi
                    }
                },
                check for zero,
            },
        },
        nodes near coords black white=420,
    ]
        \addplot[
            matrix plot,
            mesh/cols=2,
            point meta=explicit,draw=gray
        ] table [meta=C] {
            x y C
0 0 826
1 0 174
0 1 305
1 1 695
};
    \end{axis}
\end{tikzpicture}
%

%% file: figures/conf2l.tex
\begin{tikzpicture}[scale=1.0]
    \begin{axis}[
    	title={\scriptsize$\mbox{ARI} = 0.2732$},
	title style={at={(0.5,0.875)}},
        width=3.0cm,
        height=3.0cm,
    colormap={bluewhite}{color=(white) rgb255=(100,149,237)},
        xticklabels={0,1},
        xtick={0,...,1},
        xtick style={draw=none},
        yticklabels={0,1},
        ytick={0,...,1},
        ytick style={draw=none},
        enlargelimits=false,
        xlabel style={font=\scriptsize},
        ylabel style={font=\scriptsize},
        xticklabel style={font=\scriptsize},
        yticklabel style={font=\scriptsize},
        point meta min=0,
        point meta max=850,
        nodes near coords={\pgfmathprintnumber\pgfplotspointmeta},
        nodes near coords black white/.style={
            small value/.style={
                font=\scriptsize,
                yshift=-7pt,
                text=black,
                /pgf/number format/fixed,
                /pgf/number format/precision=0,
                /pgf/number format/zerofill
            },
            large value/.style={
                font=\scriptsize,
                yshift=-7pt,
                text=white,
                /pgf/number format/fixed,
                /pgf/number format/precision=0,
                /pgf/number format/zerofill
            },
            every node near coord/.style={
                check for zero/.code={
                    \pgfmathfloatifflags{\pgfplotspointmeta}{0}{
                        \pgfkeys{/tikz/coordinate}
                    }{
                        \begingroup
                        \pgfkeys{/pgf/fpu}
                        \pgfmathparse{\pgfplotspointmeta<#1}
                        \global\let\result=\pgfmathresult
                        \endgroup
                        %
                        %
                        \pgfmathfloatcreate{1}{1.0}{0}
                        \let\ONE=\pgfmathresult
                        \ifx\result\ONE
                            \pgfkeysalso{/pgfplots/small value}
                        \else
                            \pgfkeysalso{/pgfplots/large value}
                        \fi
                    }
                },
                check for zero,
            },
        },
        nodes near coords black white=420,
    ]
        \addplot[
            matrix plot,
            mesh/cols=2,
            point meta=explicit,draw=gray
        ] table [meta=C] {
            x y C
0 0 811
1 0 189
0 1 288
1 1 712
};
    \end{axis}
\end{tikzpicture}
%

%% file: figures/conf2m.tex
\begin{tikzpicture}[scale=1.0]
    \begin{axis}[
    	title={\scriptsize$\mbox{ARI} = 0.3099$},
	title style={at={(0.5,0.875)}},
        width=3.0cm,
        height=3.0cm,
    colormap={bluewhite}{color=(white) rgb255=(100,149,237)},
        xticklabels={0,1},
        xtick={0,...,1},
        xtick style={draw=none},
        yticklabels={0,1},
        ytick={0,...,1},
        ytick style={draw=none},
        enlargelimits=false,
        xlabel style={font=\scriptsize},
        ylabel style={font=\scriptsize},
        xticklabel style={font=\scriptsize},
        yticklabel style={font=\scriptsize},
        point meta min=0,
        point meta max=850,
        nodes near coords={\pgfmathprintnumber\pgfplotspointmeta},
        nodes near coords black white/.style={
            small value/.style={
                font=\scriptsize,
                yshift=-7pt,
                text=black,
                /pgf/number format/fixed,
                /pgf/number format/precision=0,
                /pgf/number format/zerofill
            },
            large value/.style={
                font=\scriptsize,
                yshift=-7pt,
                text=white,
                /pgf/number format/fixed,
                /pgf/number format/precision=0,
                /pgf/number format/zerofill
            },
            every node near coord/.style={
                check for zero/.code={
                    \pgfmathfloatifflags{\pgfplotspointmeta}{0}{
                        \pgfkeys{/tikz/coordinate}
                    }{
                        \begingroup
                        \pgfkeys{/pgf/fpu}
                        \pgfmathparse{\pgfplotspointmeta<#1}
                        \global\let\result=\pgfmathresult
                        \endgroup
                        %
                        %
                        \pgfmathfloatcreate{1}{1.0}{0}
                        \let\ONE=\pgfmathresult
                        \ifx\result\ONE
                            \pgfkeysalso{/pgfplots/small value}
                        \else
                            \pgfkeysalso{/pgfplots/large value}
                        \fi
                    }
                },
                check for zero,
            },
        },
        nodes near coords black white=420,
    ]
        \addplot[
            matrix plot,
            mesh/cols=2,
            point meta=explicit,draw=gray
        ] table [meta=C] {
            x y C
0 0 790
1 0 210
0 1 233
1 1 767
};
    \end{axis}
\end{tikzpicture}
%

%% file: figures/conf2n.tex
\begin{tikzpicture}[scale=1.0]
    \begin{axis}[
    	title={\scriptsize$\mbox{ARI} = 0.3167$},
	title style={at={(0.5,0.875)}},
        width=3.0cm,
        height=3.0cm,
    colormap={bluewhite}{color=(white) rgb255=(100,149,237)},
        xticklabels={0,1},
        xtick={0,...,1},
        xtick style={draw=none},
        yticklabels={0,1},
        ytick={0,...,1},
        ytick style={draw=none},
        enlargelimits=false,
        xlabel style={font=\scriptsize},
        ylabel style={font=\scriptsize},
        xticklabel style={font=\scriptsize},
        yticklabel style={font=\scriptsize},
        point meta min=0,
        point meta max=850,
        nodes near coords={\pgfmathprintnumber\pgfplotspointmeta},
        nodes near coords black white/.style={
            small value/.style={
                font=\scriptsize,
                yshift=-7pt,
                text=black,
                /pgf/number format/fixed,
                /pgf/number format/precision=0,
                /pgf/number format/zerofill
            },
            large value/.style={
                font=\scriptsize,
                yshift=-7pt,
                text=white,
                /pgf/number format/fixed,
                /pgf/number format/precision=0,
                /pgf/number format/zerofill
            },
            every node near coord/.style={
                check for zero/.code={
                    \pgfmathfloatifflags{\pgfplotspointmeta}{0}{
                        \pgfkeys{/tikz/coordinate}
                    }{
                        \begingroup
                        \pgfkeys{/pgf/fpu}
                        \pgfmathparse{\pgfplotspointmeta<#1}
                        \global\let\result=\pgfmathresult
                        \endgroup
                        %
                        %
                        \pgfmathfloatcreate{1}{1.0}{0}
                        \let\ONE=\pgfmathresult
                        \ifx\result\ONE
                            \pgfkeysalso{/pgfplots/small value}
                        \else
                            \pgfkeysalso{/pgfplots/large value}
                        \fi
                    }
                },
                check for zero,
            },
        },
        nodes near coords black white=420,
    ]
        \addplot[
            matrix plot,
            mesh/cols=2,
            point meta=explicit,draw=gray
        ] table [meta=C] {
            x y C
0 0 572
1 0 428
0 1 9
1 1 991
};
    \end{axis}
\end{tikzpicture}
%

%% file: figures/conf2o_cb.tex
\begin{tikzpicture}[scale=1.0]
    \begin{axis}[
    	title={\scriptsize$\mbox{ARI} = 0.3246$},
	title style={at={(0.5,0.875)}},
        width=3.0cm,
        height=3.0cm,
    colormap={bluewhite}{color=(white) rgb255=(100,149,237)},
        xticklabels={0,1},
        xtick={0,...,1},
        xtick style={draw=none},
        yticklabels={0,1},
        ytick={0,...,1},
        ytick style={draw=none},
        enlargelimits=false,
        xlabel style={font=\scriptsize},
        ylabel style={font=\scriptsize},
        xticklabel style={font=\scriptsize},
        yticklabel style={font=\scriptsize},
        colorbar,
        colorbar style={
            ytick={0,250,500,750,1000},
            yticklabels={0,250,500,750,1000},
            at={(1.1,1.0)},
            width=5pt,
            yticklabel={\pgfmathprintnumber\tick},
            yticklabel style={font=\scriptsize,
                    /pgf/number format/fixed,
            /pgf/number format/precision=0,
            /pgf/number format/1000 sep={}}
        },
        point meta min=0,
        point meta max=1000,
        nodes near coords={\pgfmathprintnumber\pgfplotspointmeta},
        nodes near coords black white/.style={
            small value/.style={
                font=\scriptsize,
                yshift=-7pt,
                text=black,
                /pgf/number format/fixed,
                /pgf/number format/precision=0,
                /pgf/number format/zerofill
            },
            large value/.style={
                font=\scriptsize,
                yshift=-7pt,
                text=white,
                /pgf/number format/fixed,
                /pgf/number format/precision=0,
                /pgf/number format/zerofill
            },
            every node near coord/.style={
                check for zero/.code={
                    \pgfmathfloatifflags{\pgfplotspointmeta}{0}{
                        \pgfkeys{/tikz/coordinate}
                    }{
                        \begingroup
                        \pgfkeys{/pgf/fpu}
                        \pgfmathparse{\pgfplotspointmeta<#1}
                        \global\let\result=\pgfmathresult
                        \endgroup
                        %
                        %
                        \pgfmathfloatcreate{1}{1.0}{0}
                        \let\ONE=\pgfmathresult
                        \ifx\result\ONE
                            \pgfkeysalso{/pgfplots/small value}
                        \else
                            \pgfkeysalso{/pgfplots/large value}
                        \fi
                    }
                },
                check for zero,
            },
        },
        nodes near coords black white=420,
    ]
        \addplot[
            matrix plot,
            mesh/cols=2,
            point meta=explicit,draw=gray
        ] table [meta=C] {
            x y C
0 0 808
1 0 192
0 1 238
1 1 762
};
    \end{axis}
\end{tikzpicture}
%

%% file: figures/heatmap1.tex
\begin{adjustbox}{width=1.0\textwidth,center}
    \pgfplotstableset{
        every head row/.style={
            typeset cell/.code={
            \ifnum\pgfplotstablecol=\pgfplotstablecols
            \pgfkeyssetvalue{/pgfplots/table/@cell content}{\rotatebox{90}{##1}\\}%
            \else
            \ifnum\pgfplotstablecol=1
            \pgfkeyssetvalue{/pgfplots/table/@cell content}{\raisebox{6pt}{\normalsize\textbf{##1}}&}%
            \else
            \pgfkeyssetvalue{/pgfplots/table/@cell content}{\rotatebox{90}{##1}&}%
            \fi
            \fi
            }
        },
    }
\pgfplotstabletypeset[%
     color cells={min=0.0,max=1.5,textcolor=black},
    /pgf/number format/fixed,
    /pgf/number format/precision=2,
    col sep=comma,
    columns/Family/.style={reset styles,string type}%
]{
Family,Adload,Agent,Alureon,BHO,CeeInject,Cycbot.G,DelfInject,FakeRean,Hotbar,Lolyda,Obfuscator,OnLineGames,Rbot,Renos,Startpage,Vobfus,Vundo,Winwebsec,Zbot,Zeroaccess
Adload,1.0,0.669,0.8817,0.0119,0.6772,0.719,0.268,0.762,0.7533,0.8798,0.8335,0.8779,0.7412,0.8519,0.769,0.0061,0.8099,0.9506,0.6414,0.7446
Agent,0.669,1.0,0.1119,0.5138,0.0416,0.0098,0.0391,0.0067,0.6576,0.1463,0.0428,0.1264,0.0591,0.0489,0.1777,0.5533,0.0236,0.0252,0.0085,0.1718
Alureon,0.8817,0.1119,1.0,0.643,0.0196,0.0613,0.2467,0.0781,0.7105,0.5927,0.0157,-0.0001,0.0117,0.0698,0.585,0.7105,0.0468,0.0322,0.0118,0.5804
BHO,0.0119,0.5138,0.643,1.0,0.6303,0.5459,0.2048,0.5533,0.3385,0.6887,0.6303,0.6592,0.5593,0.635,0.5958,0.0005,0.6303,0.7004,0.4842,0.6788
CeeInject,0.6772,0.0416,0.0196,0.6303,1.0,0.0155,0.1207,0.0113,0.7173,0.3597,0.001,0.0283,0.0018,0.0128,0.3891,0.5608,-0.0003,0.022,0.0118,0.3384
Cycbot,0.719,0.0098,0.0613,0.5459,0.0155,1.0,0.0976,0.0048,0.585,0.0224,0.0181,0.0191,0.0059,0.0062,0.2732,0.6208,0.0178,0.0012,0.0011,0.1718
DelfInject,0.268,0.0391,0.2467,0.2048,0.1207,0.0976,1.0,0.0725,0.4814,0.1969,0.1932,0.2588,0.1669,0.1769,0.0616,0.2476,0.1669,0.1292,0.0831,0.0437
FakeRean,0.762,0.0067,0.0781,0.5533,0.0113,0.0048,0.0725,1.0,0.6772,0.141,0.0203,0.0821,0.023,0.0028,0.2711,0.6129,0.0104,0.0088,0.0124,0.2168
Hotbar,0.7533,0.6576,0.7105,0.3385,0.7173,0.585,0.4814,0.6772,1.0,0.7568,0.7258,0.7088,0.6608,0.6755,0.7241,0.3167,0.7224,0.7603,0.7673,0.7568
Lolyda,0.8798,0.1463,0.5927,0.6887,0.3597,0.0224,0.1969,0.141,0.7568,1.0,0.3561,0.5804,0.3953,0.2272,0.5623,0.7241,0.2669,0.0352,0.0124,0.4786
Obfuscator,0.8335,0.0428,0.0157,0.6303,0.001,0.0181,0.1932,0.0203,0.7258,0.3561,1.0,0.035,-0.0003,-0.0003,0.4067,0.7038,0.0015,0.0226,0.0109,0.442
OnLineGames,0.8779,0.1264,-0.0001,0.6592,0.0283,0.0191,0.2588,0.0821,0.7088,0.5804,0.035,1.0,0.0106,0.0123,0.5989,0.7071,0.0472,0.0232,0.0122,0.5927
Rbot,0.7412,0.0591,0.0117,0.5593,0.0018,0.0059,0.1669,0.023,0.6608,0.3953,-0.0003,0.0106,1.0,0.0017,0.4248,0.5789,0.0027,0.0335,0.012,0.4132
Renos,0.8519,0.0489,0.0698,0.635,0.0128,0.0062,0.1769,0.0028,0.6755,0.2272,-0.0003,0.0123,0.0017,1.0,0.4209,0.6722,0.0142,0.01,0.0124,0.3979
Startpage,0.769,0.1777,0.585,0.5958,0.3891,0.2732,0.0616,0.2711,0.7241,0.5623,0.4067,0.5989,0.4248,0.4209,1.0,0.6559,0.3431,0.3853,0.0124,0.0005
Vobfus,0.0061,0.5533,0.7105,0.0005,0.5608,0.6208,0.2476,0.6129,0.3167,0.7241,0.7038,0.7071,0.5789,0.6722,0.6559,1.0,0.6559,0.7655,0.5254,0.5623
Vundo,0.8099,0.0236,0.0468,0.6303,-0.0003,0.0178,0.1669,0.0104,0.7224,0.2669,0.0015,0.0472,0.0027,0.0142,0.3431,0.6559,1.0,0.014,0.0118,0.3099
Winwebsec,0.9506,0.0252,0.0322,0.7004,0.022,0.0012,0.1292,0.0088,0.7603,0.0352,0.0226,0.0232,0.0335,0.01,0.3853,0.7655,0.014,1.0,0.0028,0.3246
Zbot,0.6414,0.0085,0.0118,0.4842,0.0118,0.0011,0.0831,0.0124,0.7673,0.0124,0.0109,0.0122,0.012,0.0124,0.0124,0.5254,0.0118,0.0028,1.0,0.0124
Zeroaccess,0.7446,0.1718,0.5804,0.6788,0.3384,0.1718,0.0437,0.2168,0.7568,0.4786,0.442,0.5927,0.4132,0.3979,0.0005,0.5623,0.3099,0.3246,0.0124,1.0
}
\end{adjustbox}

%% file: figures/bar_t.tex
\begin{tikzpicture}[scale=0.95, every node/.style={scale=1.0}]
\begin{axis}[
        width  = 0.85*\textwidth,
        height = 7.5cm,
        ymin=0,ymax=15.2,
        ytick={0,2,4,6,8,10,12},
        major x tick style = transparent,
        ybar=5*\pgflinewidth,
        bar width=10.0pt,
        ylabel = {Total ARI},
        symbolic x coords={Adload,Agent,Alureon,BHO,CeeInject,Cycbot,DelfInject,FakeRean,Hotbar,Lolyda,Obfuscator,OnLineGames,Rbot,Renos,Startpage,Vobfus,Vundo,Winwebsec,Zbot,Zeroaccess},
	y tick label style={
    		/pgf/number format/.cd,
   		fixed,
   		fixed zerofill,
    		precision=0},
        xtick = data,
        x tick label style={
        		rotate=60,
		font=\small,
		anchor=north east,
		inner sep=0mm},
        nodes near coords,
        every node near coord/.append style={rotate=90, 
        								   anchor=west,
								   font=\footnotesize,
								   /pgf/number format/.cd,
								   fixed,
								   fixed zerofill,
								   precision=2},
        enlarge x limits=0.05,
        legend cell align=left,
        legend style={
                at={(0.89,0.02)},
                anchor=south,
                column sep=1ex
        },
]
\addplot [fill=blue,opacity=1.00]
coordinates {
(Adload,12.848)
(Agent,3.4331)
(Alureon,5.4093)
(BHO,9.704)
(CeeInject,3.9589)
(Cycbot,3.1965)
(DelfInject,3.2556)
(FakeRean,3.5675)
(Hotbar,12.4961)
(Lolyda,7.4228)
(Obfuscator,4.4587)
(OnLineGames,5.3801)
(Rbot,4.1021)
(Renos,4.2483)
(Startpage,7.6574)
(Vobfus,10.1803)
(Vundo,4.095)
(Winwebsec,4.2466)
(Zbot,2.6463)
(Zeroaccess,7.2372)
};
\end{axis}
\end{tikzpicture}

%% file: figures/bar_a.tex
\begin{tikzpicture}[scale=0.95, every node/.style={scale=1.0}]
\begin{axis}[
        width  = 0.85*\textwidth,
        height = 7.5cm,
        ymin=0.0,ymax=0.775,
        ytick={0.0,0.1,0.2,0.3,0.4,0.5,0.6,0.7},
        major x tick style = transparent,
        ybar=5*\pgflinewidth,
        bar width=10.0pt,
        ylabel = {Average ARI},
        symbolic x coords={Adload,Agent,Alureon,BHO,CeeInject,Cycbot,DelfInject,FakeRean,Hotbar,Lolyda,Obfuscator,OnLineGames,Rbot,Renos,Startpage,Vobfus,Vundo,Winwebsec,Zbot,Zeroaccess},
	y tick label style={
    		/pgf/number format/.cd,
   		fixed,
   		fixed zerofill,
    		precision=1},
        xtick = data,
        x tick label style={
        		rotate=60,
		font=\small,
		anchor=north east,
		inner sep=0mm},
        nodes near coords,
        every node near coord/.append style={rotate=90, 
        								   anchor=west,
								   font=\footnotesize,
								   /pgf/number format/.cd,
								   fixed,
								   fixed zerofill,
								   precision=2},
        enlarge x limits=0.05,
        legend cell align=left,
        legend style={
                at={(0.89,0.02)},
                anchor=south,
                column sep=1ex
        },
]
\addplot [fill=blue,opacity=1.00]
coordinates {
(Adload,0.6762)
(Agent,0.1807)
(Alureon,0.2847)
(BHO,0.5107)
(CeeInject,0.2084)
(Cycbot,0.1682)
(DelfInject,0.1713)
(FakeRean,0.1878)
(Hotbar,0.6577)
(Lolyda,0.3907)
(Obfuscator,0.2347)
(OnLineGames,0.2832)
(Rbot,0.2159)
(Renos,0.2236)
(Startpage,0.403)
(Vobfus,0.5358)
(Vundo,0.2155)
(Winwebsec,0.2235)
(Zbot,0.1393)
(Zeroaccess,0.3809)
};
\coordinate (A) at (axis cs:Adload,0.5);
\coordinate (O1) at (rel axis cs:0,0);
\coordinate (O2) at (rel axis cs:1,0);
\draw [red] (A -| O1) -- (A -| O2);
\end{axis}
\end{tikzpicture}

%% file: figures/force1c.tex
    \begin{tikzpicture}[scale=1.0]
 

    \node[circle,draw,thick,color=black,fill=gray,inner sep=2.9pt] (A) at (6.85,2.225) {0};
    \node[circle,draw,thick,color=black,inner sep=2.9pt] (B) at (7.55,1.425) {1};
    \node[circle,draw,thick,color=black,inner sep=2.9pt] (C) at (6.65,2.9) {2};
    \node[circle,draw,thick,color=black,dashed,inner sep=2.9pt] (D) at (0.6,0.4) {3};
    \node[circle,draw,thick,color=black,inner sep=2.9pt] (E) at (5.35,2.0) {4};
    \node[circle,draw,thick,color=black,inner sep=2.9pt] (F) at (7.5,2.8) {5};
    \node[circle,draw,thick,color=black,dashed,inner sep=2.9pt] (G) at (7.0,0.8) {6};
    \node[circle,draw,thick,color=black,inner sep=2.9pt] (H) at (8.15,1.8) {7};
    \node[circle,draw,thick,color=black,inner sep=2.9pt] (I) at (7.9,2.9) {8};
    \node[circle,draw,thick,color=black,inner sep=2.9pt] (J) at (6.75,1.6) {9};
    \node[circle,draw,thick,color=black,inner sep=1.75pt] (K) at (6.0,3.0) {10};
    \node[circle,draw,thick,color=black,inner sep=1.75pt] (L) at (8.1,2.5) {11};
    \node[circle,draw,thick,color=black,inner sep=1.75pt] (M) at (7.1,3.1) {12};
    \node[circle,draw,thick,color=black,inner sep=1.75pt] (N) at (6.1,2.6) {13};
    \node[circle,draw,thick,color=black,inner sep=1.75pt] (O) at (7.55,1.8) {14};
    \node[circle,draw,thick,color=black,dashed,inner sep=1.75pt] (P) at (0.6,5.45) {15};
    \node[circle,draw,thick,color=black,inner sep=1.75pt] (Q) at (5.5,2.5) {16};
    \node[circle,draw,thick,color=black,inner sep=1.75pt] (R) at (6.15,1.9) {17};
    \node[circle,draw,thick,color=black,inner sep=1.75pt] (S) at (5.95,1.45) {18};
    \node[circle,draw,thick,color=black,inner sep=1.75pt] (T) at (8.3,2.2) {19};

    \draw (B) -- (A);
    \draw (C) -- (A);
    \draw[dashed] (D) -- (A);
    \draw (E) -- (A);
    \draw (F) -- (A);
    \draw[dashed] (G) -- (A);
    \draw (H) -- (A);
    \draw (I) -- (A);
    \draw (J) -- (A);
    \draw (K) -- (A);
    \draw (L) -- (A);
    \draw (M) -- (A);
    \draw (N) -- (A);
    \draw (O) -- (A);
    \draw[dashed] (P) -- (A);
    \draw (Q) -- (A);
    \draw (R) -- (A);
    \draw (S) -- (A);
    \draw (T) -- (A);
   
    \end{tikzpicture}

%% file: figures/force2c.tex
    \begin{tikzpicture}[scale=1.0]
 

    \node[circle,draw,thick,color=black,dashed,inner sep=2.9pt] (A) at  (7.15,0.4) {0};
    \node[circle,draw,thick,color=black,inner sep=2.9pt] (B) at  (8.4,4.6) {1};
    \node[circle,draw,thick,color=black,inner sep=2.9pt] (C) at  (6.9,3.8) {2};
    \node[circle,draw,thick,color=black,fill=gray,inner sep=2.9pt] (D) at  (6.85,4.55) {3};
    \node[circle,draw,thick,color=black,inner sep=2.9pt] (E) at  (5.6,4.6) {4};
    \node[circle,draw,thick,color=black,inner sep=2.9pt] (F) at  (8.15,4.15) {5};
    \node[circle,draw,thick,color=black,dashed,inner sep=2.9pt] (G) at  (6.25,3.15) {6};
    \node[circle,draw,thick,color=black,inner sep=2.9pt] (H) at  (6.9,5.2) {7};
    \node[circle,draw,thick,color=black,dashed,inner sep=2.9pt] (I) at  (5.25,4.05) {8};
    \node[circle,draw,thick,color=black,inner sep=2.9pt] (J) at  (7.5,4.2) {9};
    \node[circle,draw,thick,color=black,inner sep=1.75pt] (K) at  (6.05,5.25) {10};
    \node[circle,draw,thick,color=black,inner sep=1.75pt] (L) at  (6.0,4.25) {11};
    \node[circle,draw,thick,color=black,inner sep=1.75pt] (M) at  (7.4,5.45) {12};
    \node[circle,draw,thick,color=black,inner sep=1.75pt] (N) at  (7.9,4.8) {13};
    \node[circle,draw,thick,color=black,inner sep=1.75pt] (O) at  (6.5,5.45) {14};
    \node[circle,draw,thick,color=black,dashed,inner sep=1.75pt] (P) at  (0.6,2.0) {15};
    \node[circle,draw,thick,color=black,inner sep=1.75pt] (Q) at  (7.7,5.15) {16};
    \node[circle,draw,thick,color=black,inner sep=1.75pt] (R) at  (6.325,3.925) {17};
    \node[circle,draw,thick,color=black,dashed,inner sep=1.75pt] (S) at  (7.6,3.7) {18};
    \node[circle,draw,thick,color=black,inner sep=1.75pt] (T) at  (5.9,4.95) {19};

    \draw (B) -- (D);
    \draw (C) -- (D);
    \draw[dashed] (A) -- (D);
    \draw (E) -- (D);
    \draw (F) -- (D);
    \draw[dashed] (G) -- (D);
    \draw (H) -- (D);
    \draw[dashed] (I) -- (D);
    \draw (J) -- (D);
    \draw (K) -- (D);
    \draw (L) -- (D);
    \draw (M) -- (D);
    \draw (N) -- (D);
    \draw (O) -- (D);
    \draw[dashed] (P) -- (D);
    \draw (Q) -- (D);
    \draw (R) -- (D);
    \draw[dashed] (S) -- (D);
    \draw (T) -- (D);
   
    \end{tikzpicture}

%% file: figures/force3c.tex
    \begin{tikzpicture}[scale=1.0]
 

    \node[circle,draw,thick,color=black,inner sep=2.9pt] (A) at  (2.275,1.8) {0};
    \node[circle,draw,thick,color=black,inner sep=2.9pt] (B) at  (6.25,3.75) {1};
    \node[circle,draw,thick,color=black,inner sep=2.9pt] (C) at  (5.2,1.95) {2};
    \node[circle,draw,thick,color=black,dashed,inner sep=2.9pt] (D) at  (2.35,0.4) {3};
    \node[circle,draw,thick,color=black,inner sep=2.9pt] (E) at  (4.25,4.75) {4};
    \node[circle,draw,thick,color=black,inner sep=2.9pt] (F) at  (0.55,2.2) {5};
    \node[circle,draw,thick,color=black,dashed,inner sep=2.9pt] (G) at  (5.4,0.95) {6};
    \node[circle,draw,thick,color=black,inner sep=2.9pt] (H) at  (3.25,5.5) {7};
    \node[circle,draw,thick,color=black,fill=gray,inner sep=2.9pt] (I) at  (3.8,3.3) {8};
    \node[circle,draw,thick,color=black,inner sep=2.9pt] (J) at  (0.75,3.25) {9};
    \node[circle,draw,thick,color=black,inner sep=1.75pt] (K) at  (6.45,2.2) {10};
    \node[circle,draw,thick,color=black,inner sep=1.75pt] (L) at  (6.1,4.65) {11};
    \node[circle,draw,thick,color=black,inner sep=1.75pt] (M) at  (5.15,5.35) {12};
    \node[circle,draw,thick,color=black,inner sep=1.75pt] (N) at  (1.75,5.15) {13};
    \node[circle,draw,thick,color=black,inner sep=1.75pt] (O) at  (0.8,4.2) {14};
    \node[circle,draw,thick,color=black,dashed,inner sep=1.75pt] (P) at  (8.4,4.45) {15};
    \node[circle,draw,thick,color=black,inner sep=1.75pt] (Q) at  (6.95,3.0) {16};
    \node[circle,draw,thick,color=black,inner sep=1.75pt] (R) at  (1.9,2.8) {17};
    \node[circle,draw,thick,color=black,inner sep=1.75pt] (S) at  (2.25,4.25) {18};
    \node[circle,draw,thick,color=black,inner sep=1.75pt] (T) at  (3.75,1.7) {19};

    \draw (B) -- (I);
    \draw (C) -- (I);
    \draw[dashed] (D) -- (I);
    \draw (E) -- (I);
    \draw (F) -- (I);
    \draw[dashed] (G) -- (I);
    \draw (H) -- (I);
    \draw (A) -- (I);
    \draw (J) -- (I);
    \draw (K) -- (I);
    \draw (L) -- (I);
    \draw (M) -- (I);
    \draw (N) -- (I);
    \draw (O) -- (I);
    \draw[dashed] (P) -- (I);
    \draw (Q) -- (I);
    \draw (R) -- (I);
    \draw (S) -- (I);
    \draw (T) -- (I);
   
    \end{tikzpicture}

%% file: figures/force4c.tex
    \begin{tikzpicture}[scale=1.0]
 

    \node[circle,draw,thick,color=black,dashed,inner sep=2.9pt] (A) at  (0.5,5.5) {0};
    \node[circle,draw,thick,color=black,inner sep=2.9pt] (B) at  (8.0,4.3) {1};
    \node[circle,draw,thick,color=black,inner sep=2.9pt] (C) at  (7.4,3.9) {2};
    \node[circle,draw,thick,color=black,dashed,inner sep=2.9pt] (D) at  (2.75,0.4) {3};
    \node[circle,draw,thick,color=black,inner sep=2.9pt] (E) at  (7.9,3.8) {4};
    \node[circle,draw,thick,color=black,inner sep=2.9pt] (F) at  (6.4,3.4) {5};
    \node[circle,draw,thick,color=black,dashed,inner sep=2.9pt] (G) at  (8.35,4.9) {6};
    \node[circle,draw,thick,color=black,inner sep=2.9pt] (H) at  (5.6,4.7) {7};
    \node[circle,draw,thick,color=black,dashed,inner sep=2.9pt] (I) at  (7.1,3.0) {8};
    \node[circle,draw,thick,color=black,inner sep=2.9pt] (J) at  (6.3,4.7) {9};
    \node[circle,draw,thick,color=black,inner sep=1.75pt] (K) at  (7.45,4.45) {10};
    \node[circle,draw,thick,color=black,inner sep=1.75pt] (L) at  (5.9,4.1) {11};
    \node[circle,draw,thick,color=black,inner sep=1.75pt] (M) at  (7.275,3.5) {12};
    \node[circle,draw,thick,color=black,inner sep=1.75pt] (N) at  (5.4,4.3) {13};
    \node[circle,draw,thick,color=black,inner sep=1.75pt] (O) at  (5.6,3.75) {14};
    \node[circle,draw,thick,color=black,fill=gray,inner sep=1.75pt] (P) at  (6.675,4.25) {15};
    \node[circle,draw,thick,color=black,inner sep=1.75pt] (Q) at  (6.8,5.0) {16};
    \node[circle,draw,thick,color=black,inner sep=1.75pt] (R) at  (6.25,3.8) {17};
    \node[circle,draw,thick,color=black,inner sep=1.75pt] (S) at  (6.2,5.1) {18};
    \node[circle,draw,thick,color=black,inner sep=1.75pt] (T) at  (7.4,5.0) {19};

    \draw (B) -- (P);
    \draw (C) -- (P);
    \draw[dashed] (D) -- (P);
    \draw (E) -- (P);
    \draw (F) -- (P);
    \draw[dashed] (G) -- (P);
    \draw (H) -- (P);
    \draw[dashed] (I) -- (P);
    \draw (J) -- (P);
    \draw (K) -- (P);
    \draw (L) -- (P);
    \draw (M) -- (P);
    \draw (N) -- (P);
    \draw (O) -- (P);
    \draw[dashed] (A) -- (P);
    \draw (Q) -- (P);
    \draw (R) -- (P);
    \draw (S) -- (P);
    \draw (T) -- (P);
   
    \end{tikzpicture}

%% file: figures/conf7.tex
\begin{tikzpicture}[scale=1.0]
    \begin{axis}[
        width=8cm,
        height=8cm,
    colormap={bluewhite}{color=(white) rgb255=(100,149,237)},
        xticklabels={0,1,2,3,4,5,6},
        xtick={0,...,6},
        xtick style={draw=none},
        yticklabels={0,1,2,3,4,5,6},
        ytick={0,...,6},
        ytick style={draw=none},
        enlargelimits=false,
        xlabel={Predicted Clusters},
        ylabel={True Clusters},
        colorbar,
        colorbar style={
            ytick={0,100,200,300,400,500,600,700,800},
            yticklabels={0,100,200,300,400,500,600,700,800},
            yticklabel={\pgfmathprintnumber\tick},
            yticklabel style={
                    /pgf/number format/fixed,
            /pgf/number format/precision=0}
        },
        point meta min=0,
        point meta max=800,
        nodes near coords={\pgfmathprintnumber\pgfplotspointmeta},
        nodes near coords black white/.style={
            small value/.style={
                yshift=-7pt,
                text=black,
                /pgf/number format/fixed,
                /pgf/number format/precision=0,
                /pgf/number format/zerofill
            },
            large value/.style={
                yshift=-7pt,
                text=white,
                /pgf/number format/fixed,
                /pgf/number format/precision=0,
                /pgf/number format/zerofill
            },
            every node near coord/.style={
                check for zero/.code={
                    \pgfmathfloatifflags{\pgfplotspointmeta}{0}{
                        \pgfkeys{/tikz/coordinate}
                    }{
                        \begingroup
                        \pgfkeys{/pgf/fpu}
                        \pgfmathparse{\pgfplotspointmeta<#1}
                        \global\let\result=\pgfmathresult
                        \endgroup
                        %
                        %
                        \pgfmathfloatcreate{1}{1.0}{0}
                        \let\ONE=\pgfmathresult
                        \ifx\result\ONE
                            \pgfkeysalso{/pgfplots/small value}
                        \else
                            \pgfkeysalso{/pgfplots/large value}
                        \fi
                    }
                },
                check for zero,
            },
        },
        nodes near coords black white=400,
    ]
        \addplot[
            matrix plot,
            mesh/cols=7,
            point meta=explicit,draw=gray
        ] table [meta=C] {
            x y C
0 0 3
1 0 579
2 0 16
3 0 3
4 0 146
5 0 222
6 0 31
0 1 2
1 1 660
2 1 0
3 1 14
4 1 155
5 1 154
6 1 15
0 2 0
1 2 27
2 2 747
3 2 9
4 2 56
5 2 160
6 2 1
0 3 39
1 3 450
2 3 22
3 3 39
4 3 377
5 3 71
6 3 2
0 4 0
1 4 206
2 4 0
3 4 0
4 4 794
5 4 0
6 4 0
0 5 12
1 5 468
2 5 13
3 5 6
4 5 175
5 5 291
6 5 35
0 6 0
1 6 5
2 6 0
3 6 365
4 6 15
5 6 104
6 6 511
        };
    \end{axis}
\end{tikzpicture}
%

%% file: figures/conf4.tex
\begin{tikzpicture}[scale=1.0]
    \begin{axis}[
        width=6cm,
        height=6cm,
    colormap={bluewhite}{color=(white) rgb255=(100,149,237)},
        xticklabels={0,1,2,3},
        xtick={0,...,3},
        xtick style={draw=none},
        yticklabels={0,1,2,3},
        ytick={0,...,3},
        ytick style={draw=none},
        enlargelimits=false,
        xlabel={Predicted Clusters},
        ylabel={True Clusters},
        colorbar,
        colorbar style={
            ytick={0,175,350,525,700,875},
            yticklabels={0,175,350,525,700,875},
            yticklabel={\pgfmathprintnumber\tick},
            yticklabel style={
                    /pgf/number format/fixed,
            /pgf/number format/precision=0}
        },
        point meta min=0,
        point meta max=875,
        nodes near coords={\pgfmathprintnumber\pgfplotspointmeta},
        nodes near coords black white/.style={
            small value/.style={
                yshift=-7pt,
                text=black,
                /pgf/number format/fixed,
                /pgf/number format/precision=0,
                /pgf/number format/zerofill
            },
            large value/.style={
                yshift=-7pt,
                text=white,
                /pgf/number format/fixed,
                /pgf/number format/precision=0,
                /pgf/number format/zerofill
            },
            every node near coord/.style={
                check for zero/.code={
                    \pgfmathfloatifflags{\pgfplotspointmeta}{0}{
                        \pgfkeys{/tikz/coordinate}
                    }{
                        \begingroup
                        \pgfkeys{/pgf/fpu}
                        \pgfmathparse{\pgfplotspointmeta<#1}
                        \global\let\result=\pgfmathresult
                        \endgroup
                        %
                        %
                        \pgfmathfloatcreate{1}{1.0}{0}
                        \let\ONE=\pgfmathresult
                        \ifx\result\ONE
                            \pgfkeysalso{/pgfplots/small value}
                        \else
                            \pgfkeysalso{/pgfplots/large value}
                        \fi
                    }
                },
                check for zero,
            },
        },
        nodes near coords black white=430,
    ]
        \addplot[
            matrix plot,
            mesh/cols=4,
            point meta=explicit,draw=gray
        ] table [meta=C] {
            x y C
0 0 11
1 0 550
2 0 74
3 0 365
0 1 0
1 1 860
2 1 9
3 1 131
0 2 0
1 2 30
2 2 824
3 2 146
0 3 0
1 3 156
2 3 44
3 3 800
        };
    \end{axis}
\end{tikzpicture}
%

%% file: figures/conf3.tex
\begin{tikzpicture}[scale=1.0]
    \begin{axis}[
        width=5.5cm,
        height=5.5cm,
    colormap={bluewhite}{color=(white) rgb255=(100,149,237)},
        xticklabels={0,1,2},
        xtick={0,...,2},
        xtick style={draw=none},
        yticklabels={0,1,2},
        ytick={0,...,2},
        ytick style={draw=none},
        enlargelimits=false,
        xlabel={Predicted Clusters},
        ylabel={True Clusters},
        colorbar,
        colorbar style={
            ytick={0,155,310,465,620,775},
            yticklabels={0,155,310,465,620,775},
            yticklabel={\pgfmathprintnumber\tick},
            yticklabel style={
                    /pgf/number format/fixed,
            /pgf/number format/precision=0}
        },
        point meta min=0,
        point meta max=775,
        nodes near coords={\pgfmathprintnumber\pgfplotspointmeta},
        nodes near coords black white/.style={
            small value/.style={
                yshift=-7pt,
                text=black,
                /pgf/number format/fixed,
                /pgf/number format/precision=0,
                /pgf/number format/zerofill
            },
            large value/.style={
                yshift=-7pt,
                text=white,
                /pgf/number format/fixed,
                /pgf/number format/precision=0,
                /pgf/number format/zerofill
            },
            every node near coord/.style={
                check for zero/.code={
                    \pgfmathfloatifflags{\pgfplotspointmeta}{0}{
                        \pgfkeys{/tikz/coordinate}
                    }{
                        \begingroup
                        \pgfkeys{/pgf/fpu}
                        \pgfmathparse{\pgfplotspointmeta<#1}
                        \global\let\result=\pgfmathresult
                        \endgroup
                        %
                        %
                        \pgfmathfloatcreate{1}{1.0}{0}
                        \let\ONE=\pgfmathresult
                        \ifx\result\ONE
                            \pgfkeysalso{/pgfplots/small value}
                        \else
                            \pgfkeysalso{/pgfplots/large value}
                        \fi
                    }
                },
                check for zero,
            },
        },
        nodes near coords black white=390,
    ]
        \addplot[
            matrix plot,
            mesh/cols=3,
            point meta=explicit,draw=gray
        ] table [meta=C] {
            x y C
0 0 473
1 0 163
2 0 364
0 1 257
1 1 25
2 1 718
0 2 212
1 2 17
2 2 771
};
    \end{axis}
\end{tikzpicture}
%